# OPEN DATA, GREY DATA, AND STEWARDSHIP: UNIVERSITIES AT THE PRIVACY FRONTIER

*Christine L. Borgman*[†]

## ABSTRACT

As universities recognize the inherent value in the data they collect and hold, they encounter unforeseen challenges in stewarding those data in ways that balance accountability, transparency, and protection of privacy, academic freedom, and intellectual property. Two parallel developments in academic data collection are converging: (1) open access requirements, whereby researchers must provide access to their data as a condition of

DOI: https://doi.org/10.15779/Z38B56D489

© 2018 Christine L. Borgman.

† Distinguished Professor and Presidential Chair in Information Studies, University of California, Los Angeles. This Article is based on the Tenth Annual Berkeley Law Privacy Lecture, hosted by the Berkeley Center for Law and Technology on November 16, 2017. http://christineborgman.info

Full disclosure: The author is actively engaged in the University of California activities mentioned herein. She was a founding member of the UCLA Privacy and Data Protection Board, a member of the PISI Steering Committee, Co-Chair of the UCLA Data Governance Task Force, and currently is Chair of the University of California Academic Computing and Communications Committee (UCACC) (2017–2018 academic year; Vice Chair 2015–2017). In her role as a UCACC officer, she is a member of the UC Office of the President Cyber Risk Governance Committee (2015–2018). She has been a member of the Advisory Board to the Electronic Privacy Information Center (EPIC) since its founding in 1994 and served on the EPIC Board of Directors from 2010 to 2017. The opinions in this Article are her own.

Acknowledgements are due to the many colleagues in the University of California who have aided, abetted, and supported these privacy initiatives: Amy Blum, Jim Chalfant, Dana Cuff, Jim Davis, Jerry Kang, David Kay, Leah Lievrouw, Gene Lucas, Maryann Martone, Joanne Miller, Jan Reiff, Sheryl Vacca, Kent Wada, Scott Waugh, Shane White, and other members of the PISI and DGTF committees. My research group at the UCLA Center for Knowledge Infrastructures provided essential critique and commentary on the Article and talk: Bernadette Boscoe, Peter Darch, Milena Golshan, Irene Pasquetto, and Michael Scroggins. Morgan Wofford provided extensive bibliographic research. James Dempsey of the Berkeley Center for Law and Technology provided detailed comments on the draft paper. Outside UC, credit is due to Marc Rotenberg and the staff at the Electronic Privacy Information Center as well as Anne Washington of George Mason University. Special thanks are due to Chris Jay Hoofnagle, Paul Schwarz, and others at the Berkeley Center for Law and Technology, whose invitation to give the Tenth Annual BCLT Privacy Lecture provided the incentive to write this Article, and to Erwin Chemerinsky (Berkeley) and Katie Shilton (University of Maryland) who provided extensive and insightful commentary as respondents to the public lecture on November 16, 2017.

obtaining grant funding or publishing results in journals; and (2) the vast accumulation of "grey data" about individuals in their daily activities of research, teaching, learning, services, and administration. The boundaries between research and grey data are blurring, making it more difficult to assess the risks and responsibilities associated with any data collection. Many sets of data, both research and grey, fall outside privacy regulations such as HIPAA, FERPA, and PII. Universities are exploiting these data for research, learning analytics, faculty evaluation, strategic decisions, and other sensitive matters. Commercial entities are besieging universities with requests for access to data or for partnerships to mine them. The privacy frontier facing research universities spans open access practices, uses and misuses of data, public records requests, cyber risk, and curating data for privacy protection. This Article explores the competing values inherent in data stewardship and makes recommendations for practice by drawing on the pioneering work of the University of California in privacy and information security, data governance, and cyber risk.

# TABLE OF CONTENTS

> The world's most valuable resource is no longer oil, but data.[1]
>
> If you can't protect it, don't collect it.[2]

## I.  FRAMING THE PROBLEM

Universities are stewards of vast amounts of data. These data provide many new opportunities for research, teaching, administration, partnerships, and strategic planning. Data take many forms, have many origins, and have many uses. Data ownership is rarely clear, especially for research data, and the costs and mechanisms for stewardship are poorly understood. Although data are difficult to manage and govern in any institution, universities face a particularly complex set of responsibilities and risks.

Stewardship of data and of public trust are sometimes asymmetrical. The university community, which includes students, faculty, staff, and many other stakeholders, expects a reasonable degree of confidentiality in their dealings with an institution of research and learning. They also expect the university to respect their privacy and to keep their data secure. Furthermore, faculty and students expect their universities to respect their academic and intellectual freedom while managing and governing data. The public, which extends beyond the university community, expects universities to be fair, transparent, and accountable for resources. Good stewardship means releasing some kinds of data and preventing the release of other kinds of data. The same data may fall into either category, depending on the time, purpose, or entity requesting access. Few universities, or other institutions, have adequate governance mechanisms to address these stewardship challenges effectively.

This broad set of concerns was framed succinctly by the University of California Privacy and Information Security Initiative (PISI) which was charged in 2010 by then-President Mark Yudof to make recommendations for an overarching privacy framework to address the university's statutory and regulatory obligations; governance, implementation, and accountability structures; and policy vehicles for university policy and practice in privacy and information security.[3] As evidence of the importance placed on this effort, the President's office selected members of the PISI Steering Committee from the upper echelons of the University, including the Provost, General Counsel, Chief Compliance and Audit Officer, Chief Information Officer, and representatives from the campuses and the Academic Senate.

---

1. *The World's Most Valuable Resource Is No Longer Oil, but Data*, ECONOMIST (May 6, 2017), www.economist.com/leaders/2017/05/06/the-worlds-most-valuable-resource-is-no-longer-oil-but-data [https://perma.cc/L2HD-J7NL].

2. This is an increasingly common privacy and security aphorism. *See* Richard Bejtlich, *New Cybersecurity Mantra: "If You Can't Protect It, Don't Collect It"*, BROOKINGS INSTITUTION: TECHTANK (Sept. 3, 2015), https://www.brookings.edu/blog/techtank/2015/09/03/new-cybersecurity-mantra-if-you-cant-protect-it-dont-collect-it/ [https://perma.cc/8DDT-64RA].

In considering its charge, "the Steering Committee was guided by the following principles"[4]:

- We must maximally enable the mission of the University by supporting the values of academic and intellectual freedom.
- We must be good stewards of the information entrusted to the University.
- We must ensure that the University has access to information resources for legitimate business purposes.
- We must have a University community with clear expectations of privacy—both privileges and obligations of individuals and of the institution.
- We must make decisions within an institutional context.
- We must acknowledge the distributed nature of information stewardship at UC, where responsibility for privacy and information security resides at every level.

These principles have proved to be robust in the several years since the final report was submitted to the President and the Regents. Most of the recommendations have been adopted and implemented, including appointing Chief Privacy Officers and establishing joint Academic Senate-Administration boards on privacy and information security on each of the ten campuses. At the UC-wide level, the Academic Senate monitors PISI implementation via the UC Academic Computing and Communications Committee (UCACC). Individual campuses have extended the PISI principles in various ways. UCLA, which established a joint Senate-Administration Privacy and Data Protection Board in 2005, extended the PISI principles and recommendations in the Data Governance Task Force Report.[5]

---

3. UC PRIVACY & INFO. SEC. INITIATIVE STEERING COMM., PRIVACY AND INFORMATION SECURITY INITIATIVE STEERING COMMITTEE REPORT TO THE PRESIDENT 1, 27–28 (2013), http://ucop.edu/privacy-initiative/uc-privacy-and-information-security-steering-committee-final-report.pdf [https://perma.cc/6ANW-HCCK].

4. *Id.* at 1–2.

5. UCLA DATA GOVERNANCE TASK FORCE, UCLA DATA GOVERNANCE TASK FORCE FINAL REPORT AND RECOMMENDATIONS 1, 8 (2016), http://evc.ucla.edu/reports/DGTF-report.pdf [https://perma.cc/97TQ-3AGM].

Identifying problems and principles is an essential starting point to address the challenges of the day. Applying these principles to solve these problems is much harder. Over the last several years, the complexity of these challenges has become ever more apparent. This Article explores the current landscape of opportunities, responsibilities, risks, and frontiers facing universities in a data-rich world. It draws on the pioneering work of the University of California, one of the world's premier public research universities, at the forefront of both data governance and data exploitation. It also draws on a large body of work on policy and practice for governing access to research data.

The epigraphs at the top of the Article frame the arguments herein. Data have become the "new oil" as one of the modern world's most valued commodities. Market leaders, whether in commerce or in higher education, may be those most adept at exploiting data in their realms. As non-consumptive goods, arguably more valuable than the finite supply of oil, data can be mined, combined, and reused for multiple applications over long periods of time.[6] The aphorism, "if you can't protect it, don't collect it," has circulated in the privacy, security, and hacker communities for a decade or more. Leaking data can be at least as dangerous as leaking oil. For universities to sustain the public trust, and to live by the principles that guided the UC Privacy and Security Initiative, they must address the converse of that aphorism: "if you collect it, you must protect it."

## II. THE DATA-RICH WORLD OF RESEARCH UNIVERSITIES

Universities are stewards of many kinds of data, some of which they collect, others that they acquire, and yet others that are byproducts of regular activities. The value of these data, the possibilities for exploitation, the responsibilities for stewardship, and the types of associated risks vary immensely.

Intentional data collection is the more obvious sort, such as materials gathered by investigators as part of research projects and information about current and prospective students gathered by the registrar. These data tend to be governed by established mechanisms such as grant contracts, Institutional Review Boards, HIPAA, and FERPA.[7] At the other extreme is incidental data collection that is difficult to identify or govern, such as that gathered by students, by staff in administrative roles, and by technology such as security cameras controlled within offices or departments. A growing source of incidental data collection is software packages that individuals install on university networks for otherwise legitimate purposes in teaching and research. In between is a vast array of data collection that may be more or less intentional, more or less governed, and whose applications may be more or less foreseeable at the time of collection. These include learning management systems, personnel systems that include faculty dossiers for academic evaluation and promotion, identity cards that encode various privileges (library usage, food service, building access, debit charges, etc.), and much more.

---

6. *See* CHRISTINE L. BORGMAN, BIG DATA, LITTLE DATA, NO DATA: SCHOLARSHIP IN THE NETWORKED WORLD 7 (2015); *see also generally* CHARLOTTE HESS & ELINOR OSTROM, UNDERSTANDING KNOWLEDGE AS A COMMONS: FROM THEORY TO PRACTICE (2007).

7. *See, e.g.,* Health Insurance Portability and Accountability Act of 1996, Pub. L. No. 104-191, 110 Stat. 1936; Family Educational Rights and Privacy Act of 1974, 20 U.S.C. § 1232g (2012); Protection of Human Subjects, 45 C.F.R § 46 (2009).

In all of these arenas, data volumes and variety are growing at rates far greater than most administrators or faculty realize. Those individuals who recognize the value and opportunities in these data are not necessarily obligated to seek permission to exploit them. Third parties outside the university may be the first to recognize data opportunities, and approach individuals at any level of the university for partnerships. Governance mechanisms to assure protection of privacy, academic freedom, intellectual property, information security, and compliance with regulations in the uses of such data are nascent, at best.

Of this immense landscape of data issues in universities, this Article focuses on two exemplars. The first is research data, spanning all academic domains from the sciences, technology, and medicine to the social sciences, humanities, and the arts. Although the data management issues in these areas are critical and far from solved, at least two decades of practice and policy inform current discussions. The second is data collected by universities about members of its community, including students, faculty, staff, visitors, patients, and other stakeholders. Data collection about individual persons, both intentional and incidental, is accelerating rapidly with the implementation of systems that can exploit "data exhaust" from the activities of individuals.[8] Despite several decades of research on principles and practice for "privacy by design," developers too often default to collecting as much data as possible.[9]

---

8.   VIKTOR MAYER-SCHÖNBERGER & KENNETH CUKIER, BIG DATA: A REVOLUTION THAT WILL TRANSFORM HOW WE LIVE, WORK, AND THINK 113 (2013) ("Data exhaust . . . refers to data that is shed as a byproduct of people's actions and movements in the world . . . Many companies design their systems so that they can harvest data exhaust and recycle it, to improve an existing service or to develop new ones.").

9.   *See generally* Victoria Bellotti & Abigail Sellen, *Design for Privacy in Ubiquitous Computing Environments, in* PROCEEDINGS OF THE THIRD EUROPEAN CONFERENCE ON COMPUTER-SUPPORTED COOPERATIVE WORK 77 (Giorgio de Michelis, Carla Simone, & Kjeld Schmidt eds., 1993); Herbert Burkert, *Privacy-Enhancing Technologies: Typology, Critique, Vision, in*

Data that universities collect about their communities is also a large and diverse category. The primary exemplar discussed herein is teaching and student learning, itself an area of data explosion. Fully online courses can capture data on every keystroke of every participant, if they choose to do so, creating rich profiles on individual students and interactions between them. Less obvious is the amount of data produced in hybrid courses, where learning management systems complement interactions in campus classrooms. Students acquire their readings and assignments online, participate in online discussions and other activities, and submit their assignments through these systems, all of which is discretely time-stamped. When universities aggregate this learning data with other kinds of data they hold on their students, extensive profiles result. These datasets can be deployed for learning analytics, institutional reports to government and accreditation agencies, academic research, or for surveillance of activities and behavior.

## A.    RESEARCH DATA

Scholars collected research data long before the advent of the scholarly journal, which is barely 350 years old.[10] Data are reported in publications, usually in selected and synthesized forms. Some data are kept for reuse by investigators; other data may be bartered in exchange for still further data or as invitations to collaborate.[11] Until recently, data were considered part of the research process, rather than products to be disseminated. Data release has become a condition of obtaining grants and publishing papers in many domains, especially in the biosciences and medicine.[12] Survey research in the

---

TECHNOLOGY AND PRIVACY: THE NEW LANDSCAPE 125 (Philip E. Agre & Marc Rotenberg eds., 1997); Ian Goldberg, David Wagner & Eric Brewer, *Privacy-Enhancing Technologies for the Internet*, *in* PROCEEDINGS OF THE IEEE COMPCON '97 at 103 (1997); Katie Shilton, *Participatory Personal Data: An Emerging Research Challenge for the Information Sciences*, 63 J. AM. SOC'Y FOR INFO. SCI. TECH. 1905 (2012); Katie Shilton, *Values Levers: Building Ethics into Design*, 38 SCI. TECH. & HUMAN VALUES 374 (2012).

10. *350 Years of Scientific Publishing*, ROYAL SOC'Y, https://royalsociety.org/journals/publishing-activities/publishing350/ [https://perma.cc/7RZ9-BLRG] (last visited June 15, 2018).

11. Stephen Hilgartner & Sherry I. Brandt-Rauf, *Data Access, Ownership, and Control: Toward Empirical Studies of Access Practices*, 15 KNOWLEDGE: CREATION DIFFUSION UTILIZATION 355, 357, 363–66 (1994).

12. *See, e.g.*, Joseph S. Ross & Harlan M. Krumholz, *Ushering in a New Era of Open Science Through Data Sharing: The Wall Must Come Down*, 309 JAMA 1355, 1356 (2013); Joseph S. Ross, *Clinical Research Data Sharing: What an Open Science World Means for Researchers Involved in Evidence Synthesis*, 5 SYSTEMATIC REVS. 159 at 1 (2016); Geoffrey Boulton et al., *Science as a Public Enterprise: The Case for Open Data*, 377 LANCET 1633, 1634 (2011).

social sciences has a long history of data sharing. For example, in the humanities, archaeology is a growth area for data sharing and archiving.[13]

When datasets were small and locally controlled, issues of data stewardship and governance rarely arose. As datasets grew larger and distributed collaborations became more common, tools to mine and combine data became more sophisticated. These opportunities vary immensely between domains, universities, countries, and cultures, as do applicable policies.[14] As the volume of publicly available research data expands, concerns for stewardship of these data become more urgent.[15]

*1. Scope and Definitions*

One part of the challenge in managing research data is the difficulty of defining "research" or "data" succinctly. Information, documents, and materials exist in many forms and in many states, only a portion of which might be considered research data for the purposes of governance. The Oxford English Dictionary is a good starting place to define concepts such as research:

> The act of searching carefully for or pursuing a specified thing or person; an instance of this. Systematic investigation or inquiry aimed at contributing to knowledge of a theory, topic, etc., by careful consideration, observation, or study of a subject. In later use also: original critical or scientific investigation carried out under the auspices of an academic or other institution. Investigation undertaken in order to obtain material for a book, article, thesis, etc.; an instance of this.

Locating a singular definition of research used within the University of California proved similarly elusive. At UCLA, for example, the Office of Research Administration lists responsibilities and resources on its website but does not define research in its glossary of terms. Research, like beauty, is often

---

13.  *See generally* Eric Kansa, *Openness and Archaeology's Information Ecosystem*, 44 WORLD ARCHAEOLOGY 498 (2012); Eric C. Kansa, Sarah Whitcher Kansa & Benjamin Arbuckle, *Publishing and Pushing: Mixing Models for Communicating Research Data in Archaeology*, 9 INT'L J. DIG. CURATION 57 (2014).

14.  BORGMAN, *supra* note 6, at 55–58.

15.  *See, e.g.*, Francine Berman & Vint Cerf, *Who Will Pay for Public Access to Research Data?*, 341 SCIENCE 616 (2013); Tony Hey & Anne E. Trefethen, *Cyberinfrastructure for e-Science*, 308 SCIENCE 817 (2005); Jeremy York, Myron Gutmann & Francine Berman, What Do We Know About the Stewardship Gap? 1 (July 17, 2016) (unpublished manuscript), https://deepblue.lib.umich.edu/bitstream/handle/ 2027.42/122726/StewardshipGap_Final.pdf [https://perma.cc/26V3-VQXT].

in the eye of the beholder, who may be a grant-funding program manager or an academic personnel officer.[16]

One area in which firm definitions are needed are studies involving human subjects. In the United States, such studies fall under the regulation of the federal Department of Health and Human Services. DHHS regulations define research as "a systematic investigation, including research development, testing and evaluation, designed to develop or contribute to generalizable knowledge."[17] If a study meets these requirements and is deemed to involve human subjects, then the protocol must be submitted to the Institutional Review Board (IRB) of the university. Whether a study is considered research or involves human subjects is not always obvious. A systematic study that involves a survey of students for the purposes of university strategic planning is usually not considered research because it is not intended for publication, and thus not for generalizable knowledge. Relatedly, systematic investigations of human activity that are intended for publication, but that do not require direct contact with individual living persons, may or may not be deemed research for the purpose of IRB review. Further, problems arise when data collected for administrative purposes later are deemed worthy of publication, which is not an uncommon occurrence.

"Research data" is similarly problematic to define and is often left undefined in guidelines for releasing or depositing data from a research project. At best, data may be defined by example, such as observations, facts, samples, or records. A definition developed elsewhere is the basis for this Article's discussion: "data refers to entities used as evidence of phenomena for the purposes of research or scholarship."[18] This phenomenological definition covers data in any academic discipline, recognizing that one scholar's signal is another's noise.

### 2. Open Access to Research Data

Practices and policies for open access to research data are intertwined with those for open access to scholarly publications such as journal articles. Since the early days of "electronic publishing" in the 1990s, activists have called for open access to scholarly publications as a means to democratize access to information.[19] Open access has taken many forms, such as disseminating

---

16. *See* DONALD E. STOKES, PASTEUR'S QUADRANT: BASIC SCIENCE AND TECHNOLOGICAL INNOVATION 16 (1997).

17. 45 C.F.R. § 46.102(d) (2017).

18. BORGMAN, *supra* note 6, at 29.

19. *See* Stevan Harnad, *The PostGutenberg Galaxy: How to Get There from Here*, 11 INFO. SOC'Y 285, 288 (1995) ("[T]he general public, which is likewise gaining greater access to the Net, also stands to benefit from the free availability of scholarly literature . . . ."); Stevan preprints prior to publication, disseminating post-prints after publication, or publishing in journals that are free to read online.[20] Scholars are also publishing a growing number of books in open access formats, often with print-on-demand options. Open access increases the dissemination of research, which tends to enhance the visibility of authors and their institutions, so the payoffs are several. Economic models for open access dissemination vary widely, as do stewardship models; and responsibility for access and for sustainability often fall to different parties.[21]

Providing access to research data is often a condition for publishing an article, whether or not the article itself is published in an open access form.[22] Thus, data release usually occurs at the time of submitting a paper for publication. Datasets can be contributed to archives or repositories, which assign them a unique identification number, and that ID is linked to the paper. Ideally, it becomes possible to search for data and identify associated publications, or to search for publications and identify associated datasets.[23] Publishing articles in open access venues and disseminating preprints are more established practices than is providing open access to data. Data release varies widely by domain, with the greatest acceptance in the biosciences and medicine, and by type of data, research method, funding source, and other factors.[24]

---

Harnad, *Post-Gutenberg Galaxy: The Fourth Revolution in the Means of Production of Knowledge*, 2 PUBLIC-ACCESS COMPUTER SYS. REV. 39, 47 (1991) ("A decade and half later my own rewarding experience with electronic skywriting has convinced me that this newest medium's unique potential to support and sustain open peer commentary must now be made generally available too . . . .").

20.   *See* PETER SUBER, OPEN ACCESS 97–98 (2012).

21.   *See, e.g.*, Isabel Bernal, *Open Access and the Changing Landscape of Research Impact Indicators: New Roles for Repositories*, 1 PUBLICATIONS 56, 58–60 (2013); CHRISTINE L. BORGMAN, SCHOLARSHIP IN THE DIGITAL AGE: INFORMATION, INFRASTRUCTURE, AND THE INTERNET 255, 259–60 (2007); *Open Access Policies*, HARVARD UNIV., https://osc.hul.harvard.edu/policies [https://perma.cc/S5G9-SHUY]; Jennifer Howard, *Open Access Gains Major Support in U. of California's Systemwide Move*, CHRON. HIGHER EDUC. (Aug. 2, 2013), www.chronicle.com/article/Open-Access-Gains-Major/140851 [https://perma.cc/3KB7-RW42]; *UC Open Access Policies*, OFFICE OF SCHOLARLY COMMUNICATION, http://osc.universityofcalifornia.edu/open-access-policy/ [https://perma.cc/T9KD-QAEX]; Richard Van Noorden, *Europe Joins UK Open-Access Bid*, 487 NATURE 285, 285 (2012). *See also generally* JOHN WILLINSKY, THE ACCESS PRINCIPLE: THE CASE FOR OPEN ACCESS TO RESEARCH AND SCHOLARSHIP (2006); Randall Munroe, *The Rise of Open Access*, 342 SCIENCE 58 (2013).

22.   *See* BORGMAN, *supra* note 6, at 48; GEOFFREY BOULTON ET AL., SCIENCE AS AN OPEN ENTERPRISE 27 (2012) https://royalsociety.org/~/media/policy/projects/sape/2012-06-20-saoe.pdf [https://perma.cc/2APE-E2BW].

23.   BORGMAN, *supra* note 21, at 116–18; *see* Philip E. Bourne, *Will a Biological Database Be Different from a Biological Journal?*, 1 PLOS COMPUTATIONAL BIOLOGY 179, 180–81 (2005).

24.   *See generally* BORGMAN, *supra* note 21; *see also* BORGMAN, *supra* note 6, at 260–64.

A legacy of open access publishing that contributes to stewardship challenges is that the notion of "publication" has become more diffuse. Whether something can be considered a formal publication matters for evaluating scholarship, and thus for hiring, tenure, grant proposals, library collections, and much more. In the print world, publications were more readily distinguishable from "grey literature." The latter category consists of documents such as working papers, reports, pamphlets, and preprints that have scholarly value, but that have not been vetted by peer review or disseminated through a formal publication process. In the online world, versions of scholarly documents proliferate. The same or similar content, often with the same or similar titles and authors, may appear as preprints, post-prints, working papers, slide decks, and as the formal "official" version of a publication. Initial versions of documents may or may not become formal publications at a later time. Others may diverge into multiple publications. Choosing which version to cite is a judgment call by the citing author.

The publication versioning problem intersects with the data stewardship problem in at least two ways. One is determining the relationship between a dataset and a publication or other document that describes the dataset. Research projects can generate many versions of publications and many versions of datasets, resulting in a complex array of many-to-many relationships between datasets and publications explaining the context in which they were created.

The second problem is the differing degrees of validation and of permanence of publications and datasets. The popular term "data publishing" suggests that data and publications are released to the scholarly communication system with comparable status.[25] Similarly, data citation is promoted as a means to encourage data release by giving comparable scholarly credit.[26] This equivalence is also embedded in technology by assigning Digital Object Identifiers (DOI) to each article and dataset.[27] DOIs are a formal system of persistent and unique identifiers that is managed by scholarly publishers, libraries, and other stakeholders.[28] However, journal articles are subject to far more scrutiny, and to greater investments in stewardship, than are datasets.[29] In scholarly communication, publishing implies a process of peer review, validation, dissemination, and access.[30] Publishers and libraries provide stewardship and access.

---

25.  *See* Mark A. Parsons & Peter A. Fox, *Is Data Publication the Right Metaphor?*, 12 DATA SCI. J. WDS32, WDS40 (2013); BORGMAN, *supra* note 6, at 225–27.

26.  *See generally* NAT'L RESEARCH COUNCIL, FOR ATTRIBUTION—DEVELOPING DATA ATTRIBUTION AND CITATION PRACTICES AND STANDARDS: SUMMARY OF AN INTERNATIONAL WORKSHOP 210 (Paul F. Uhlir ed., 2012); CODATA-ICSTI Task Grp. on Data Citation Standards & Practices, *Out of Cite, Out of Mind: The Current State of Practice, Policy, and Technology for the Citation of Data*, 12 DATA SCI. J. CIDCR1, CIDCR14–CIDCR15 [hereinafter CODATA-ICSTI Task Group].

27.  *See* CODATA-ICSTI Task Group, *supra* note 26, at CIDCR32; NAT'L RESEARCH COUNCIL, *supra* note 26, at 52.

In contrast, datasets are published only in the dictionary sense of "making public."[31] Rarely are datasets peer-reviewed. Although data repositories may assess datasets for technical standards, such as adequate metadata and documentation, responsibility for scholarly or scientific quality is left to the contributors.[32] Long-term accessibility of datasets is a significant concern. Datasets may remain available only for fixed time periods at the end of a grant project and funding for repositories is often unstable. When datasets are released by posting on local websites, links degrade quickly.[33] Authors may assume, all too readily, that assigning a DOI to an object, whether a journal article, conference paper, dataset, presentation slide deck, glass slide, or other entity, makes that item a publication and assures long-term accessibility. Unfortunately, assigning a persistent identifier only ensures that the item has a unique ID. It does not guarantee that the ID will retrieve any content.[34]

---

28. *See, e.g., Discussion Board for the Persistent Identifiers Working Group RDA*, RESEARCH DATA ALLIANCE, https://www.rd-alliance.org/groups/pid-interest-group.html [https://perma.cc/YKX9-82VS] (last visited Feb. 20, 2018) ("The purpose of the Persistent Identifier Interest Group is to synchronize identifier-related efforts, address important and emerging PID-related topics and coordinate activities, including appropriate RDA Working Groups, to practically solve PID-related issues from the engaged communities."); Jan Brase, Michael Lautenschlager & Irina Sens, *The Tenth Anniversary of Assigning DOI Names to Scientific Data and a Five Year History of DataCite*, 21 D-LIB MAGAZINE (2015); *Metadata Enables Connections*, CROSSREF (Aug. 4, 2018), https://www.crossref.org/services/ [https://perma.cc/UC4A-VMQ8] (describing use of metadata to persistently catalogue and track scholarly publications); Micah Altman & Mercè Crosas, *The Evolution of Data Citation: From Principles to Implementation*, 37 IASSIST Q. 62, 65 (2013) ("Global persistent identifiers, such as DOIs and Handles, offer a mechanism to provide a permanent link that can be configured to always resolve to a web page from which the data can be accessed, independent of whether the location of that page changes over time."); *see also* Matthew S. Mayernik & Keith E. Maull, *Assessing the Uptake of Persistent Identifiers by Research Infrastructure Users*, 12 PLOS ONE 1, 1 (2017) (evaluating whether research infrastructures are being increasingly identified and referenced in the research literature to via persistent citable identifiers); Tobias Weigel et al., *A Framework for Extended Persistent Identification of Scientific Assets*, 12 DATA SCI. J. 10, 13 (2013) (presenting a framework for persistent identification that fundamentally supports context information).

29. *See* Christine L. Borgman, *Data Citation as a Bibliometric Oxymoron, in* THEORIES OF INFORMETRICS AND SCHOLARLY COMMUNICATION 93, 94 (Cassidy R. Sugimoto ed., 2015) ("Scholarly publication normally requires peer review and dissemination in a venue with recognized status for credit and attribution. Journals and books usually meet this standard of publication. . . . Data are far more complex objects—if they are objects at all—than the entities to which bibliometrics applies.") (internal citation omitted).

30. *See* BORGMAN, *supra* note 21, at 58–60, 65–68.

31. *See* BORGMAN, *supra* note 6, at 47–49.

32. *See generally* LOUISE CORTI ET AL., MANAGING AND SHARING RESEARCH DATA: A GUIDE TO GOOD PRACTICE (1st ed. 2014) (outlining a comprehensive set of best practices for data management and sharing); *see also* Veerle Van den Eynden & Louise Corti, *Advancing Research Data Publishing Practices for the Social Sciences: From Archive Activity to Empowering Researchers*, 18 INT'L J. DIGITAL LIBR. 113, 119–20 (2017).

### 3. *Opportunities in Research Data*

Democratizing access to knowledge is among the drivers of open access to publications and to research data. The opportunities in these categories differ in important ways, however. Open access to publications expands readership to audiences far beyond the privileged communities that enjoy access to expensive journals and books through their university libraries. Whether read in the form of preprint, post-print, or published journal article, open access dissemination of scholarly work has created a vast international audience of interested students, researchers, enthusiasts, patients, parents, and other parties. Having the domain knowledge and linguistic ability to exploit these materials is another matter, but providing access is a good start on equity issues.

Similarly, open access to research data expands scholarly data resources far beyond the investigators who collected and analyzed them. Others can exploit these data, as intact datasets or in combination with other resources, for many purposes. The barriers of requisite domain knowledge and linguistic skills still apply, but the opportunities to exploit data are potentially boundless. Among the policy drivers commonly cited for open access are transparency, to allow others to inspect and evaluate findings; reproducibility, to verify findings by repeating a study; and reuse, whether as an independent dataset or aggregated with other data. Accountability to taxpayers, in the case of public funding, is also mentioned frequently.[35]

---

33.  *See* Borgman, *supra* note 29, at 100; Alberto Pepe et al., *How Do Astronomers Share Data? Reliability and Persistence of Datasets Linked in AAS Publications and a Qualitative Study of Data Practices among US Astronomers*, 9 PLOS ONE 1, 2 (2014); Christine L. Borgman, Andrea Scharnhorst & Milena S. Golshan, Digital Data Archives as Knowledge Infrastructures: Mediators of Data Sharing and Reuse 1 (Feb. 2, 2018) (unpublished manuscript), https://arxiv.org/abs/1802.02689 [https://perma.cc/S4AQ-SY7Z]; NAT'L SCI. BOARD, LONG-LIVED DIGITAL DATA COLLECTIONS: ENABLING RESEARCH AND EDUCATION IN THE 21ST CENTURY 34 (2005) ("Tracking is a challenge because links to the data in publications, Web sites, etc. may become obsolete. Finding the data that were previously available may be difficult for those outside the immediate project team.").

34.  *See* BORGMAN, *supra* note 6, at 258 ("In practice, neither identity nor persistence is absolute. People change names, documents change versions, digital objects change locations when transferred from one computer to another, and they change in form when migrated over generations of software . . . ."); Borgman, *supra* note 29 (explaining that persistent identifiers must be able to "resolve to a location" online to facilitate access in addition to identification).

The promises of open access to research data are vast, although mired in hyperbole. Vast stores of research data are predicted to accumulate through open access policies enforced by publishers, funding agencies, and government directives, as well as through voluntary participation. These stores can be mined and combined by anyone, at least in principle, leading to new research findings, new innovations, new companies, and new market sectors.[36] A European policy presentation at a recent Research Data Alliance meeting

35. Press Release, European Comm'n., Scientific Information in the Digital Age: Ensuring Current and Future Access for Research and Innovation (Feb. 15, 2007), https://ec.europa.eu/digital-single-market/en/news/scientific-information-digital-age-ensuring-current-and-future-access-research-and-innovation [https://perma.cc/9M7N-B3A5]; BORGMAN, *supra* note 21, at 77 ("Funding agencies are using the opportunities afforded by online access to reaffirm their responsibility to taxpayers . . . ."); Geoffrey Boulton, *Open Your Minds and Share Your Results*, 486 NATURE 441, 441 (2012) ("[A]bove all, we need scientists to accept that publicly funded research is a public resource."); BOULTON ET AL., *supra* note 22, 39 ("Access to data . . . is important for citizens' involvement in science and their pursuit of scholarship through data which, after all, for publicly funded science they have paid for through their taxes."); ORG. FOR ECON. COOPERATION & DEV., OECD PRINCIPLES AND GUIDELINES FOR ACCESS TO RESEARCH DATA FROM PUBLIC FUNDING 21–22 (2007), www.oecd.org/dataoecd/9/61/38500813.pdf [https://perma.cc/VS68-V32B]; John P. Holdren, Memorandum for the Heads of Executive Departments and Agencies: Increasing Access to the Results of Federally Funded Scientific Research 1 (Feb. 22, 2013), https://obamawhitehouse.archives.gov/sites/default/files/microsites/ostp/ostp_public_access_memo_2013.pdf [https://perma.cc/6SDZ-B3RB].

36. *See, e.g.*, Chris Anderson, *The End of Theory: The Data Deluge Makes the Scientific Method Obsolete*, WIRED (June 23, 2008, 12:00 PM), https://www.wired.com/2008/06/pb-theory/ [https://perma.cc/G2UL-C3JF] (explaining the relation between massive data and applied mathematics); Peter Arzberger et al., *An International Framework to Promote Access to Data*, 303 SCIENCE 1777, 1777 (2004) ("Open access . . . provides greater returns from the public investment in research, generates wealth through downstream commercialization of outputs, and provides decision-makers with facts needed to address complex, often transnational, problems."); Geoffrey Boulton et al., *Open Data in a Big Data World: An International Accord*, INT'L SCI. COUNCIL (2015), https://council.science/cms/2017/04/open-data-in-big-data-world_long.pdf [https://perma.cc/DHP2-QHZF]; Kenneth Cukier, *Data, Data Everywhere*, ECONOMIST (Feb. 25, 2010), http://www.economist.com/node/15557443 [https://perma.cc/9TA6-JPGB]; Dawn Field et al., *'Omics Data Sharing*, 326 SCIENCE 234, 234–35 (2009); Brooks Hanson, Andrew Sugden & Bruce Alberts, *Making Data Maximally Available*, 331 SCIENCE 649 (2011); Holdren, *supra* note 35; Scott D. Kahn, *On the Future of Genomic Data*, 331 SCIENCE 728, 728 (2011); John Palfrey & Jonathan Zittrain, *Better Data for a Better Internet*, 334 SCIENCE 1210, 1210 (2011); O. J. Reichman, Matthew B. Jones & Mark P. Schildhauer, *Challenges and Opportunities of Open Data in Ecology*, 331 SCIENCE 703, 703–04 (2011); Global Alliance for Genomics & Health, *A Federated Ecosystem for Sharing Genomic, Clinical Data*, 352 SCIENCE 1278, 1278–79 (2016).

suggested that "[b]y 2020, the European Data Economy in the most favourable scenario could contribute up to 4% of EU GDP."[37]

In practice, however, considerable investment is required to make research data useful to anyone beyond the original data collectors. Whereas most scholarly documents can be read and understood as independent units, the same is not true of data. A dataset alone, without accompanying documentation of the research methods by which it was created, analysis and interpretation of the findings, and associated context such as instruments, models, and software, may be little more than a string of numbers. The better documented and curated, the more useful any given set of data will be to others.[38]

## B.    GREY DATA: ACADEMIC, ADMINISTRATIVE, AND INSTRUCTIONAL

"Grey data" is proposed as an umbrella term to describe the vast array of data that universities accumulate outside the research realm. Analogous to grey literature, explained above, these are useful data that have not been vetted by peer review, or perhaps by any other governance mechanism of the university. Grey data have become critical to a university's ability to "innovate, enhance, and execute [its] core missions of education, research, and service."[39] Some of these data are collected for mandatory reporting obligations such as enrollments, diversity, budgets, grants, and library collections. Many types of data about individuals are collected for operational and design purposes, whether for instruction, libraries, travel, health, or student services. Universities are increasingly aware of the asset value of data about their communities. Some of these data have legal encumbrances for compliance purposes, but many are collected for reasons of internal management and external competitiveness. Outside entities also see the value in these data, whether through explicit partnerships with universities to exploit data, or by collecting data on users of their products.

The drivers of data collection in universities are many, not the least of which is "market-based solutions" as a response to the lack of funding for public colleges and universities. Higher education reform is being defined in "highly economistic terms" leading to "measurement panic."[40] University administrators may be given statistical benchmark targets for enrollments, time to degree, retention, diversity, and other countable factors, not unlike performance targets in private business. When higher education is viewed more as a job track than as an investment in a democratic citizenry, market-driven measurement may be an inevitable result. Competition looms everywhere.

---

37. CELINA RAMJOUÉ, BUILDING A EUROPEAN DATA ECONOMY: THE ROLE OF RESEARCH DATA 6 (2017).

38. *See* BORGMAN, *supra* note 6, at 4, 48; Irene V. Pasquetto, Bernadette M. Randles & Christine L. Borgman, *On the Reuse of Scientific Data*, 16 DATA SCI. J. 1, 4 (2017) ("[T]he dataset is of little value without associated documentation, and often software, code, and associated scientific models.").

39. UCLA DATA GOVERNANCE TASK FORCE, *supra* note 5, at 3.

40. Sanford F. Schram, *The Future of Higher Education and American Democracy: Introduction*, 36 NEW POL. SCI. 425, 427–30 (2014).

### 1. Collecting Grey Data

Universities always have collected data about their communities, their operations, and their services—as do businesses, governments, and public service sectors. As daily activities of teaching, learning, research, and operations have moved online, the "volume, velocity, and variety" of data collection have exploded.[41] The uses of digital data from online networks differ from those of data collected offline in at least two respects. One is that discrete data elements become far more valuable when combined with other data. Information gathered about student performance in a single course, once aggregated with data on performance in other courses, test scores, social media activity, library usage, and dietary habits, for example, yield rich profiles on individuals. The other difference between offline and online collection is that many more people have access to online data. In the past, an individual instructor knew little about students enrolled in her course beyond the list provided by the registrar. Now the instructor may be given profiles on each student to track progress. Academic counselors, student advising staff, instructional designers, registrars, department chairs, deans, provosts, and many others may also have access to these data.

The pervasiveness of information technologies has accelerated over the course of several decades, much of which originated in university environments. Today's senior faculty have lived through eras of mainframe computers, minicomputers, desktop personal computers, and ubiquitous mobile devices such as laptops, tablets, and smart phones. They have adapted their research and teaching practices to accommodate, if not to incorporate these technologies. Instrumentation large and small is deeply embedded in the practice of many domains, ranging from space telescopes to sensor networks to nanotech devices. In the early days of portable technologies, instruction practices excluded these devices from the classroom, asking students to leave their calculators, cell phones, and laptops at home, or at least out of sight. Although some faculty continue to bar mobile technologies from classrooms, most have embraced tools such as learning management systems (LMS) that support course websites, links to reading materials, discussion groups, and authentication to library and enrollment services. Pedagogy has shifted rapidly over the last decade from rejecting or ignoring students' uses of information technologies to embracing "cyberlearning," both for the analytical data generated and for the ability to adapt instruction to students' behavior.[42]

---

41. Doug Laney, *3D Data Management: Controlling Data Volume, Velocity and Variety*, META GROUP RES. NOTE (Feb. 6, 2001), https://blogs.gartner.com/doug-laney/files/2012/01/ad949-3D-Data-Management-Controlling-Data-Volume-Velocity-and-Variety.pdf [https://perma.cc/ENC3-AUBU].

### 2. Opportunities in Grey Data

As cited above, data have become the new "oil" that drives commerce and competition.[43] Google, Amazon, Facebook, and many other companies have built financial empires by collecting and combining personal data. These data are used to profile individuals, segment the population into discrete units, and present information highly selectively. They can also be used to monitor or predict behavior, resulting in closer observation for illicit or suspicious activities, or for auspicious moments to present advertisements, news, or other content. Many decisions are made about people on the basis of their online traces.

Universities, often with commercial partners, are exploiting data about individuals in similar ways. By collecting detailed data on individual student performance, some universities are creating an individualized "learning path" for each student, with various benchmarks toward degree completion.[44] Other institutions are constructing profiles that assign students to one of three categories that predict success, such as the green, yellow, and red "Stoplight" system. Some profiles incorporate data from social networks to assess a student's social connectedness algorithmically.[45] These profiles may be used to

---

make decisions about students, often without their knowledge, about jobs, scholarships, financial aid, choice of majors, counseling, and other matters.

Integrating data from multiple sources and systems is a nontrivial matter for reasons of technology, measurement, and inference.[46] The higher education community, via an EDUCAUSE initiative funded by the Gates Foundation, has proposed a "Next Generation Digital Learning Environment" that will provide greater interoperability and a freer flow of data between applications that gather data about students.[47]

## III.    UNIVERSITY RESPONSIBILITIES FOR DATA

The massive data collection by universities creates vast opportunities for research, teaching, learning, service, outreach, and strategic management. These data collections expose universities to new risks and create responsibilities that may converge and diverge in unexpected ways. Four categories of responsibilities are outlined here: stewardship and governance, and protecting privacy, academic freedom, and intellectual property. As a means to focus this vast territory, the discussion draws out issues that are common to research data and to grey data. Because privacy concerns are central to this Article, academic freedom and intellectual property are discussed in privacy contexts.

### A.    STEWARDSHIP AND GOVERNANCE

By collecting data, institutions assume responsibility for managing those data in the short and long term. Among the many descriptions of these roles, such as sustainability, curation, access, and preservation, "stewardship" has become the overarching term. Although "stewardship" is used in nuanced ways in the scientific, library, archival, and policy communities, stewardship encompasses a commitment to managing data in ways that they remain findable, accessible, and useful.[48] For some kinds of data, stewardship requires

---

46. *See generally* Franke Kreuter & Roger D. Peng, *Extracting Information from Big Data: Issues of Measurement, Inference, and Linkage*, *in* PRIVACY, BIG DATA, AND THE PUBLIC GOOD: FRAMEWORKS FOR ENGAGEMENT 257 (Julia Lane et al. eds., 1 ed. 2014).

47. *See* MALCOLM BROWN, JOANNE DEHONEY & NANCY MILLICHAP, THE NEXT GENERATION DIGITAL LEARNING ENVIRONMENT: A REPORT ON RESEARCH 3–4 (2015), https://library.educause.edu/~/media/files/library/2015/4/eli3035-pdf.pdf [https://perma.cc/9FQM-C46U].

48. *See, e.g.,* Mark D. Wilkinson et al., *The FAIR Guiding Principles for scientific data management and stewardship*, 3 SCI. DATA (2016); BROWN ET AL., *supra* note 47, at 4; Ge Peng et al., *A Unified Framework for Measuring Stewardship Practices Applied to Digital Environmental Datasets*, 13 DATA SCIENCE JOURNAL 231, 234–36 (2015); *About the National Digital Stewardship Alliance*, NAT'L DIG. STEWARDSHIP ALL., https://ndsa.org/about/ [https://perma.cc/GQG5-W4F-J] (last visited Aug. 15, 2018); Myron P. Gutmann et al., *Stewardship Gap Project*, indefinite preservation; for others, stewardship requires regular cycles of record disposal.[49] However, given the dynamic nature of these data collections, traditional archival approaches to sustaining access to static resources are unlikely to suffice. In an "age of algorithms" where datasets are in constant flux and can be disaggregated and reaggregated continuously for multiple analytical purposes, new approaches are sorely needed.[50]

Although universities have broad responsibilities for stewarding the data they collect, acquire, and hold, some individual persons, offices, committees, or other entities must take specific actions, make investments, and manage the daily operations of data stewardship. Determining which entities have which responsibilities, based on what criteria and policies, is the process of governance. The UC Privacy and Information Security Initiative (PISI), discussed in framing this Article, was among the first to address this process in U.S. higher education. The PISI principles explicitly acknowledge the "distributed nature of information stewardship at UC, where responsibility for privacy and information security resides at every level."[51] Universities are unlikely to appoint "data czars" responsible for all manner of research and grey data. More feasible is for an office or committee to wrangle generalized policies, agreements, and governance mechanisms.

*1. Research Data*

Responsibility for research data in universities generally defaults to the researchers who collected those data. These researchers have a vested interest in exploiting and protecting these data. They also are the people who know most about the data's content and context. Local knowledge is essential to data management, given the vast array of data types, domain expertise, policies, and practices. Along with the benefits of local control come limitations in expertise and continuity. In domains with external funding, graduate students and post-doctoral fellows conduct most data collection and perform most of the

---

http://www.colorado.edu/ibs/cupc/stewardship_gap/ [https://perma.cc/SC6X-CZSW] (last visited Aug. 15, 2018); *see also generally* NAT'L ACAD. SCI., ENSURING THE INTEGRITY, ACCESSIBILITY AND STEWARDSHIP OF RESEARCH DATA IN THE DIGITAL AGE (2009).

49. *See Records Retention & Disposition Guidelines*, UCLA CORP. FIN. SERS., https://www.finance.ucla.edu/tax-records/records-management/records-retention-disposition-guidelines [https://perma.cc/7AF6-D8MN] (last visited Aug. 15, 2018); Special Section on Selection, Appraisal, and Retention of Digital Scientific Data, 3 DATA SCIENCE JOURNAL 191–232 (2004); CAROL BLUM, COUNCIL ON GOVERNMENTAL RELATIONS, ACCESS TO, SHARING AND RETENTION OF RESEARCH DATA: RIGHTS & RESPONSIBILITIES (2012), https://www.cogr.edu/sites/default/files/access_to_sharing_and_retention_of_research_data-_rights_%26_responsibilities.pdf [https://perma.cc/4AFR-CBVB].

50. *See* Clifford Lynch, *Stewardship in the "Age of Algorithms"*, 22 FIRST MONDAY (2017).

51. UC PRIVACY & INFO. SEC. INITIATIVE STEERING COMM., *supra* note 3, at 8.

management tasks. Students and post-docs often write software code, scripts, and algorithms to analyze those data. Although experts in a research domain, students and post-docs rarely are also experts in data management or software engineering. They perform essential research tasks but are short-term employees who are replaced every few years as students graduate, fellowships end, and grant projects are completed.[52]

As papers are submitted for publication and grant closure looms, many authors and investigators are responsible for releasing associated data. If so, they need to find (and often to fund) ways of sustaining access to their data for some specified number of years after the granting period. The preferred solution is usually to deposit datasets in a data archive or repository, whether organized by discipline, data type, or institution, as these entities tend to have long-term commitments and staff responsible for curation. Archiving of digital research data has been under way for at least fifty years by entities such as the World Data Systems,[53] IQSS,[54] and ICPSR.[55] Some agencies fund research and data archives to sustain access to findings, such as the National Institutes of Health (U.S.) and Economic and Social Research Council (U.K.). Other funding agencies may require universities to maintain their own data archives as a condition of receiving grants.[56] Many public archives, however, are funded by research grants, which limits their ability to make indefinite commitments.

### 2. Grey Data

Responsibility for grey data is highly diffuse in universities. Those who collect data may become the stewards of those data or may pass them to other stewards inside or outside the institution. Among the many data collectors and stewards of grey data are libraries, registrars, undergraduate and graduate divisions, schools and departments, instructional development, individual faculty and staff, and administrators of housing, food services, student stores, and many more. Here too, students and other limited-term staff may have substantial responsibility for day-to-day data collection and management. Many of these data have transient value, but many may be kept indefinitely, whether for potential later use as stores cumulate or because it is often easier to keep them than to invest the labor necessary to discard records selectively.

---

52. *See* BORGMAN, *supra* note 6.

53. *See Trusted Data Services for Global Science*, ICSU WORLD DATA SYS., https://www.icsu-wds.org/ [https://perma.cc/6KBA-BHYU] (last visited Aug. 15, 2018).

54. *See* HARVARD INST. FOR QUANTITATIVE SOC. SCI., https://www.iq.harvard.edu/home [https://perma.cc/TR2X-W4TZ] (last visited Aug. 15, 2018); HARVARD DATAVERSE, https://dataverse.harvard.edu/ [https://perma.cc/6SS3-Z9N9] (last visited Aug. 15, 2018).

55. *See* INTER-UNIVERSITY CONSORTIUM FOR POLITICAL AND SOCIAL RESEARCH (ICPSR), http://www.icpsr.umich.edu/icpsrweb/ICPSR/ [https://perma.cc/9WZC-C4PV] (last visited Aug. 15, 2018).

56. *Expectations*, U.K. ENG'G & PHYSICAL SCI. RESEARCH COUNCIL, www.epsrc.ac.uk/about/standards/researchdata/expectations/ [https://perma.cc/MHQ4-5J8U] (last visited Aug. 15, 2018).

Where compliance rules for data protection and management clearly apply, universities will implement those rules. The larger problem is the growing collections of grey data where few rules are explicitly applicable and data stewards must exercise discretion.

## B.    PRIVACY

Privacy is an essential but elusive concept, as Chemerinsky,[57] Solove,[58] Nissenbaum,[59] and others have eloquently explained. It lacks a single core essence and is best understood as a pluralistic construct that spans information collection, processing, dissemination, accessibility, autonomy, and certain types of intrusion. Privacy is best understood in a context, such as a university's relationship to the data it collects, acquires, and holds. Somewhat different considerations apply to research and to grey data, although even this boundary is porous and mutable.

Privacy issues associated with data usually involve records collected about individuals—a foundational area of privacy law and policy. The Code of Fair Information Practice, known as FIPS (or FIPPS for Fair Information Practice Principles), generally applies, regardless of the intended purpose for data collection. FIPS was formulated in the early days of digital records and incorporated in the foundational U.S. laws about government data collection in the 1970s. The U.S. FIPS became the basis for the OECD principles in 1980, updated in 2013, which are widely promulgated and adopted.[60] HIPAA (medical patient records) and FERPA (educational records), for example, incorporate most of the FIPS principles.

---

57.  *See* Erwin Chemerinsky, *Rediscovering Brandeis's Right to Privacy*, 45 BRANDEIS L.J. 643, 644–45 (2006).

58.  *See* DANIEL J. SOLOVE, UNDERSTANDING PRIVACY 1–11 (2010 ed.).

59.  *See* HELEN NISSENBAUM, PRIVACY IN CONTEXT: TECHNOLOGY, POLICY, AND THE INTEGRITY OF SOCIAL LIFE 2–4 (1st ed. 2009).

60.  *See* ORG. FOR ECON. COOPERATION & DEV., THE OECD PRIVACY FRAMEWORK 69–71 (2013), http://www.oecd.org/internet/ieconomy/privacy-guidelines.htm [https://perma.cc/526E-365M] [hereinafter OECD PRIVACY FRAMEWORK]; ORG. FOR ECON. COOPERATION & DEV., OECD GUIDELINES ON THE PROTECTION OF PRIVACY AND TRANSBORDER FLOWS OF PERSONAL DATA (1980), http://www.oecd.org/sti/ieconomy /oecdguidelinesontheprotectionofprivacyandtransborderflowsofpersonaldata.htm [https://perma.cc/9GZC-PS8H] [hereinafter OECD 1980 GUIDELINES]; OFFICE OF THE ASSISTANT SEC'Y FOR PLANNING & EVALUATION, U.S. DEP'T OF HEALTH & HUMAN SERVS. RECORDS, COMPUTERS AND THE RIGHTS OF CITIZENS (1973), https://aspe.hhs.gov/report/ records-computers-and-rights-citizens [https://perma.cc/CN94-XMZW].

Requirements for notice of data collection and consent to acquire specific kinds of data are the most widely implemented of the FIPS principles. These two principles continue to be required not only in research contexts, but in credit, housing, social media, and any online service that collects data about individuals—even if the notice and consent contract is buried in the fine print of "click through" agreements.[61] Other OECD FIPS principles provide important privacy guidance, such as the Data Quality Principle, which says "personal data should be relevant to the purposes for which they are to be used, and . . . should be accurate, complete, and kept up-to-date"; the Purpose Specification Principle, which requires that the intended uses for collection be specified in advance; and the Use Limitation Principle, that subsequent uses should be limited to those specified and not repurposed without consent of the data subject, unless by other legal authority.[62] Other FIPS principles include security safeguards, openness, individual participation, and accountability.[63]

Protecting privacy by maintaining confidentiality is among the central concerns in human subjects research. Rules for the treatment of human subjects were developed in the same era as the FIPS principles. The Belmont Report, a FIPS-era foundational document, established three premises for protection of human subjects: respect for persons, beneficence, and justice. The Belmont principles, in turn, are the basis for Institutional Review Boards (IRB) at universities and other research institutions, which are administered with U.S. government oversight.[64]

Investigators who conduct human subjects research intentionally, as in much of the social sciences, health, and medical domains, submit their research proposals and protocols to the appropriate Institutional Review Board. The IRB determines whether the study complies with federal regulations and the amount of oversight required. Some studies are exempt, while others require

---

extensive and continuing review.[65] If human subjects data are to be released upon publication or conclusion of the study, de-identification and anonymization of individuals normally is required, following protocols for best practice in a given domain.

Despite the long history of privacy regulations and best practices in universities, many privacy issues are emerging in areas not clearly covered by FIPS, Institutional Review Boards, or regulations such as HIPAA, FERPA, and PII (Personally Identifiable Information). These include research projects that capture records of human activity, whether traces of online or offline activity, historical records, or incidental observations of individuals with technologies such as cameras, audio recorders, drones, or other sensors during investigations for other purposes.

Learning analytics are a primary example of grey data that contains sensitive and often personally identifiable data about individuals, but that is not subject to IRB rules for confidentiality and data protection. Some universities insist on explicit notice and consent to collect data about students' online behavior, but many assume that students have given implicit consent by enrolling in the university. Students may not know what is being collected about them, much less what is being done with those data or who has access to them.[66] FERPA provides little guidance in using or protecting these data, as learning analytics appear to fall in the generally allowable category of educational uses.[67]

The UC Privacy and Information Security Initiative and the UCLA Data Governance Task Force both addressed data privacy issues by distinguishing between two types of privacy and the security necessary to protect them, as illustrated in Figure 1.

---

65. *See* NAT'L RESEARCH COUNCIL, PROPOSED REVISIONS TO THE COMMON RULE: PERSPECTIVES OF SOCIAL AND BEHAVIORAL SCIENTISTS, WORKSHOP SUMMARY 26–27 (Robert Pool ed. 2013).

66. *See generally* Brown, *supra* note 44; Lisa Ho, *Naked in the Garden: Privacy and the Next Generation Digital Learning Environment*, EDUCAUSE REV. (July 31, 2017), https://er.educause.edu:443/articles/2017/7/naked-in-the-garden-privacy-and-the-next-generation-digital-learning-environment [https://perma.cc/9PRN-X3K4] (last visited Sep 27, 2017); *Asilomar II: Student Data and Records in the Digital Era*, STANFORD, https://sites.stanford.edu/asilomar/ [https://perma.cc/FF7G-XYPN] (last visited Aug. 15, 2018); *The Asilomar Convention for Learning Research in Higher Education*, ASILOMAR CONVENTION (June 11, 2014), http://asilomar-highered.info/asilomar-convention-20140612.pdf [https://perma.cc/E5XC-AN2L]; Selinger, *supra* note 45; Sharon Slade & Paul Prinsloo, *Learning Analytics: Ethical Issues and Dilemmas*, 57 AM. BEHAVIORAL SCIENTIST 1509 (2013).

67. *See* Steven J. McDonald, *A Few Things about E-FERPA*, EDUCAUSE REV. (Jan. 28, 2013), https://er.educause.edu:443/blogs/2013/1/a-few-things-about-eferpa [https://perma.cc/K5RC-DVJK]; Diana Orrick, *An Examination of Online Privacy Issues for Students of American Universities*, in INTERNATIONAL CONFERENCE ON INTERNET COMPUTING 330 (2003), https://www.educause.edu/ir/library/pdf/CSD4039.pdf [https://perma.cc/4E9L-937M].

Figure 1: Relationships Between Autonomy Privacy, Information Privacy, and Information Security[68]

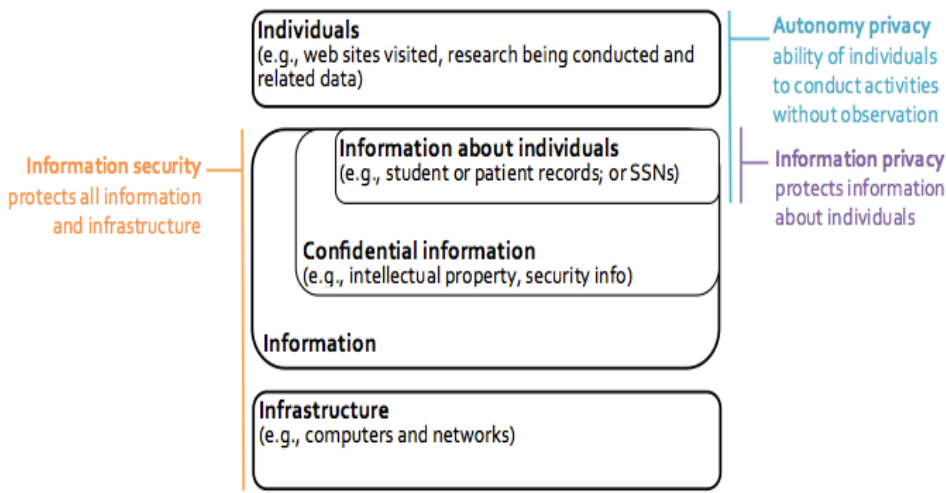

Information privacy is narrowly drawn to include specific information about individuals, such as those elements in the legal definitions of PII. In the California law, PII includes a specific list of data elements,[69] whereas the U.S. federal code is more general: "PII means information that can be used to distinguish or trace an individual's identity, either alone or when combined with other personal or identifying information that is linked or linkable to a specific individual."[70]

Autonomy privacy, or the ability of individuals to conduct activities without observation, is a larger category that subsumes PII. It includes safeguards from surveillance and other kinds of monitoring of behavior. Autonomy privacy overlaps with academic freedom concerns, as discussed in the next Section, because it includes the ability to conduct research without being observed. Information security, the third category in the PISI and DGTF reports, protects the confidentiality, integrity, and availability of information, and thus includes the protection of intellectual property. These committees were comprised of multiple, and sometimes competing, stakeholders. Both committees debated their issues for many months to reach consensus. Separating the concepts of autonomy privacy, information privacy, and information security led the committees to a broader framing of privacy, security, and governance and to more concise recommendations. These categories are loosely based on legal distinctions between informational and autonomy privacy; security is necessary because current technologies have led to an "unprecedented ability to learn the most intimate and personal things about individuals . . . [and] unprecedented access to information about individuals."[71]

## C.      ACADEMIC FREEDOM

Like privacy, academic freedom is a complex and elusive concept. In considering university responsibilities for data, it intersects with privacy and with freedom of speech. The most succinct, and most widely adopted, statement of academic freedom is that "[t]eachers are entitled to full freedom in research and in the publication of the results"[72] because academic freedom is "fundamental to the advancement of truth."[73] It is not an absolute right to free speech; rather, the formal statement of academic freedom distinguishes between speech on one's area of expertise and speech as a private citizen, and includes conditions such as adequate performance of other academic duties.[74]

Protecting autonomy privacy is essential to protecting academic freedom. In research contexts, faculty need to be able to protect research in progress, including research data, in the free pursuit of inquiry. Scholars often "test ideas in extreme form" as a means to develop hypotheses, brainstorm with collaborators, or provoke internal debate.[75] Releasing private communications risks mischaracterizing the research and the individuals involved, and thus limits the free pursuit of truth and inquiry. Research data are part of the research process, and thus similarly subject to protection on the grounds of academic freedom and autonomy privacy.[76]

Academic freedom protection is normally associated with academic tenure.[77] Autonomy privacy, however, applies more broadly to the university

---

71.  *See* Chemerinsky, *supra* note 57, at 656.

72.  *1940 Statement of Principles on Academic Freedom and Tenure*, AM. ASS'N OF UNIV. PROFESSORS 14 (1940), https://www.aaup.org/file/1940%20Statement.pdf [https://perma.cc/Y76E-P923].

73.  *See id.*

74.  *See id.*

75.  *Statement on the Principles of Scholarly Research and Public Records Requests*, UCLA JOINT SENATE-ADMIN. TASK FORCE ON ACAD. FREEDOM (July 2012), https://apo.ucla.edu/policies-forms/academic-freedom [https://perma.cc/Z3EZ-4ZBN].

76.  UCLA DATA GOVERNANCE TASK FORCE, *supra* note 5, at 3; UC PRIVACY & INFO. SEC. INITIATIVE STEERING COMM., *supra* note 3, at 2.

77.  *See* AM. ASS'N OF UNIV. PROFESSORS, *supra* note 72; *see also generally* Erwin Chemerinsky, *Is Tenure Necessary to Protect Academic Freedom?*, 41 AM. BEHAV. SCIENTIST 638 community. Non-tenured faculty, research staff, and students also conduct research and those data deserve similar protections. Autonomy privacy goes beyond the scope of academic freedom, which covers research and teaching, to include "the ability of individuals to conduct activities without observation . . . ."[78] These recent UC initiatives reinforce long-standing university policy on protecting electronic communications and media: "[t]he University recognizes that principles of academic freedom and shared governance, freedom of speech, and privacy hold important implications for the use of electronic communications."[79] With very limited exceptions, "[t]he University does not examine or disclose electronic communications records without the holder's consent."[80] These policies are strong protections against electronic surveillance. They also reinforce FIPS by requiring notice and consent to collect data about individuals.

Grey data, such as digital records about teaching and student learning, are similarly covered under the UC Electronic Communications Policy and the adopted recommendations about privacy, information security, and data governance. However, these policies appear to provide much stronger protections of privacy and academic freedom than are typical of U.S. institutions of higher education.

## D.    INTELLECTUAL PROPERTY

Making data open rests on the assumption that a data owner has the rights to release data and to grant reuse by others. Therein lies the rub. Data ownership in the realm of academic research rarely is made explicit, at least until disputes arise. Control of data often rests on agreements among collaborators, which may or may not be spelled out in grant proposals or publications.[81]

---

(1998) (addressing how the First Amendment protects the academic freedom of faculty and if there are alternatives that might provide equal safeguards).

78.   UCLA DATA GOVERNANCE TASK FORCE, *supra* note 5, at 11; UC PRIVACY & INFO. SEC. INITIATIVE STEERING COMM., *supra* note 3, at 3.

79.   *Electronic Communications Policy*, UNIV. OF CAL., OFFICE OF THE PRESIDENT 10 (2005), https://policy.ucop.edu/doc/7000470/ElectronicCommunications [https://perma.cc/T5L6-WMPC].

80.   *Id.*

81.   *See* BORGMAN, *supra* note 6, at 12, 84 (describing different approaches to data control among collaborators); *see also* Jillian Claire Wallis, The Distribution of Data Management Responsibility Within Scientific Research Groups (June 18, 2012) (unpublished Ph.D. dissertation, UCLA) https://papers.ssrn.com/sol3/papers.cfm?abstract_id=2269079 [https://perma.cc/5EB5-S8KV] (examining data management tasks performed by members of six research groups and members' perception of data management responsibilities); PAUL A. DAVID & MICHAEL SPENCE, TOWARDS INSTITUTIONAL INFRASTRUCTURES FOR E-SCIENCE: THE SCOPE OF THE CHALLENGE 92 (2003) (articulating the nature and significance

Ownership of intellectual property carries a large set of rights and responsibilities, some which are associated with privacy protection and intrusion. Corporate owners of scholarly publishing, mass media, and social media content deploy "digital rights management" (DRM) technologies to track uses and users in minute detail.[82] These technologies have eroded traditional protections of privacy and intellectual freedom in libraries and other domains.[83] Universities, hospitals, and private businesses who own or control medical patient records are responsible for protecting the confidentiality of those records and limiting their dissemination. Despite regulations, the health industry has found ways to monetize these records, thus invading privacy and causing other harms to patients.[84] Universities have special responsibilities for managing their intellectual property in ways that protect the privacy of their communities and minimize harm.

Funding agencies usually hold principal investigators responsible for data management plans and other rules associated with intellectual products of research.[85] Journals hold authors responsible for releasing or depositing data when such rules apply.[86] Scholars acquire many kinds of data over the course

---

of non-technological issues that bear on infrastructures created to enable collaborations in e-Science); Paul A. David, *Can "Open Science" Be Protected from the Evolving Regime of IPR Protections?*, 160 J. INSTITUTIONAL & THEORETICAL ECON. 9 (2004) (explaining that in some fields, legal instituion innovations are undermining sharing of access to raw data-steams and documented database resources).

82. Julie E. Cohen, *DRM and Privacy*, 18 BERKELEY TECH. L.J. 575 (2003).

83. *See generally* Julie E. Cohen, *A Right to Read Anonymously: A Closer Look at "Copyright Management" in Cyberspace*, 28 CONN. L. REV. 981 (1996) (discussing digital monitoring of individual reading habits for purposes of so-called "copyright management" in cyberspace); Clifford Lynch, *The Rise of Reading Analytics and the Emerging Calculus of Reader Privacy in the Digital World*, 22 FIRST MONDAY (2017) (illustrating the way reader privacy concerns are shifting from government to commercial surveillance and the interactions between government and the private sector).

84. *See* BEN GOLDACRE, BAD PHARMA: HOW DRUG COMPANIES MISLEAD DOCTORS AND HARM PATIENTS 283 (2012); Patrick Radden Keefe, *The Family That Built an Empire of Pain*, NEW YORKER (Oct. 30, 2017), https://www.newyorker.com/magazine/2017/10/30/the-family-that-built-an-empire-of-pain [https://perma.cc/9XNY-L9YX] (examining a specific case of two doctors whose "ruthless marketing of painkillers" lead to millions of addicted patients); James F. Peltz & Melody Petersen, *L.A. Billionaire Cancer Doctor Patrick Soon-Shiong Battles Business Turbulence*, L.A. TIMES (July 5, 2017, 3:00 AM), www.latimes.com/business/la-fi-soon-shiong-20170705-story.html [https://perma.cc/ME58-MK5H].

85. *See* Brian Westra, *Developing Data Management Services for Researchers at the University of Oregon*, *in* RESEARCH DATA MANAGEMENT: PRACTICAL STRATEGIES FOR INFORMATION PROFESSIONALS 375, 389 (Joyce M Ray ed., 2014).

86. COUNCIL OF SCI. EDITORS, CSE'S WHITE PAPER ON PROMOTING INTEGRITY IN SCIENTIFIC JOURNAL PUBLICATIONS 21–31 (2012), https://www.councilscienceeditors.org/wp-content/uploads/CSE-White-Paper_2018-update-050618.pdf [https://perma.cc/78SP-TZTY].

of their careers, often at great personal expense. As a consequence of these practices, faculty tend to hold research records, observations, physical samples, and other types of research data as their own property for most intents and purposes. For example, laboratory notebooks have special status in fields where patent protection may arise.[87] Strictly speaking, research data may be considered factual and thus not subject to copyright or to ownership.[88] However, the nature of "facts" is a subject of dispute among historians, philosophers, social scientists, and lawyers alike.[89]

Although many universities, including the University of California, claim ownership of research data, researchers may be largely unaware of these regulations unless disputes arise, or an individual faculty member wishes to take a substantial trove of data to another university when changing jobs.[90] Little guidance exists for how data ownership policies apply to data release requirements. For example, the UC policy cited for data ownership is the last sentence of this paragraph in the Academic Personnel Manual:

> 5. Publicity of Results
>
> All such research shall be conducted so as to be as generally useful as possible. To this end, the right of publication is reserved by the University. The University may itself publish the material or may authorize, in any specific case, a member or members of the faculty to publish it through some recognized scientific or professional medium of publication. A report detailing the essential data and presenting the final results must be filed with the University. **Notebooks and other original records of the research are the property of the University.**[91]

---

87. *See generally* Colin L. Bird, Cerys Willoughby & Jeremy G. Frey, *Laboratory Notebooks in the Digital Era: The Role of ELNs in Record Keeping for Chemistry and Other Sciences*, 42 CHEMICAL SOC'Y REV. 8157 (2013) (examining the foundations of the emerging opportunities for preserving and curating electronic records, focusing on ELNs); *see also* Jason T. Nickla & Matthew B. Boehm, *Proper Laboratory Notebook Practices: Protecting Your Intellectual Property*, 6 J. NEUROIMMUNE PHARMACOLOGY 4 (2011) (arguing that there is a need for research institutions to develop strict policies on the proper use and storage of research documentation); Kalpana Shankar, *Order from Chaos: The Poetics and Pragmatics of Scientific Recordkeeping*, 58 J. AM. SOC'Y FOR INF. SCI. 1457 (2007) (focusing on the process by which scientific records are created to reflect both personal need and professional needs).

88. *See* PETER BALDWIN, THE COPYRIGHT WARS: THREE CENTURIES OF TRANS-ATLANTIC BATTLE 318–83 (2014).

89. *See generally* ANN M. BLAIR, TOO MUCH TO KNOW: MANAGING SCHOLARLY INFORMATION BEFORE THE MODERN AGE (2010); Daniel Rosenberg, *Data Before the Fact, in* "RAW DATA" IS AN OXYMORON 15 (Lisa Gitelman ed., 2013) (sketching the early history of the concept of "data" in order to understand the way in which the space was formed).

90. *See, e.g.*, Bradley J. Fikes, *UC San Diego Sues USC and Scientist, Alleging Conspiracy to Take Funding, Data*, L.A. TIMES (July 5, 2015, 5:55 PM), http://www.latimes.com/local/education/la-me-ucsd-lawsuit-20150706-story.html [https://perma.cc/V88D-MTKT ] (describing how the University of California, San Diego sued the University of Southern California and a nationally recognized Alzheimer's disease researcher alleging that they conspired to take federal funding, data and employees from a U.C. San Diego study center); Larry Gordon, Gary Robbins & Bradley J. Fikes, *What's Behind UCSD, USC Court Battle?*, SAN DIEGO UNION TRIB. (July 9, 2015, 9:15 AM), http://www.sandiegouniontribune.com/news/science/sdut-usc-ucsd-alzheimers-paul-aisen-court-legal-2015jul19-story.html [https://perma.cc/7VDM-CHRW]; Gary Robbins, *UC San Diego Wins Legal Battle in Dispute with USC Over Alzheimer's Project*, L.A. TIMES (July 24, 2015, 10:34 PM), http://www.latimes.com/local/california/la-me-0725-uc-sandiego-20150725-story.html [https://perma.cc/9V59-D4QL]; Gary Robbins & Bradley J. Fikes, *USC Siphons Away Most of Alzheimer's Program*, SAN DIEGO UNION TRIB. (Aug. 29, 2015, 12:45 PM), http://www.sandiegouniontribune.com/news/science/sdut-ucsd-usc-alzheimers-aisen-cooperative-study-2015aug29-htmlstory.html [https://perma.cc/Z3Z5-GVNY] (describing how a court ruled that USC could take over a prestigious Alzheimer's disease research program long run by U.C. San Diego after a researcher left with its data).

Given the advances in research practice and digital records since the policy was established in 1958, these issues are receiving renewed attention by the Academic Senate and other UC bodies. Open access, data governance, and data ownership are among the agenda items for the UC Academic Computing and Communications Committee,[92] UC Committee on Libraries and Scholarly Communication,[93] and UC Committee on Research Policy,[94] for example.

Ownership and responsibility for grey data is particularly problematic. Although university records presumably are property of the university, many individuals and units may be involved in data collection, analysis, reporting, and management. As records are mined and combined, tracking sources and policies associated with individual datasets becomes more difficult. In principle, students own the intellectual property in their coursework, such as papers and assignments, yet some of that work and associated online activities may be captured by learning management systems or other educational technologies. When commercial partners are involved in data collection, either via university contracts or software tools deployed by individual faculty, licensing and ownership of grey data may be unclear or opaque.[95]

---

91. ROBERT G. SPROUL, UNIVERSITY OF CALIFORNIA REGULATION NO. 4 (GENERAL UNIVERSITY POLICY REGARDING ACADEMIC APPOINTEES: SPECIAL SERVICES TO INDIVIDUALS AND ORGANIZATIONS) 3 (1958) (emphasis added), http://www.ucop.edu/academic-personnel-programs/_files/apm/apm-020.pdf [https://perma.cc/M3WA-9ZWC].

92. *University Committee on Academic Computing and Communications (UCACC)*, UNIV. OF CAL. ACAD. SENATE, http://senate.universityofcalifornia.edu/committees/ucacc/index.html [https://perma.cc/VJ8K-WTAJ] (last visited Aug. 15, 2018).

93. *University Committee on Library and Scholarly Communication (UCOLASC)*, UNIV. OF CAL. ACAD. SENATE, http://senate.universityofcalifornia.edu/committees/ucolasc/index.html [https://perma.cc/39Q7-GB34] (last visited Aug. 15, 2018).

94. *University Committee on Research Policy (UCORP)*, UNIV. OF CAL. ACAD. SENATE, http://senate.universityofcalifornia.edu/committees/ucorp/index.html [https://perma.cc/GZP6-S3Q3] (last visited Aug. 15, 2018).

95. *See* UCLA DATA GOVERNANCE TASK FORCE, *supra* note 5.

## IV.    THE PRIVACY FRONTIER

The drive to collect data at ever greater volumes, velocity, and variety is moving universities into unknown territory—the "privacy frontier"—at a far faster rate than most administrators, faculty, researchers, or students are aware. Universities are competitive institutions, both internally and externally. Those who exploit data most effectively will gain research grants, awards, students, administrative efficiencies, and other rewards. Those who govern and steward their data most effectively are likely to gain greater long-term advantages. On shorter horizons, it is all too easy to exploit data in ways that risk violations of privacy. Protecting privacy adds a layer of complexity to exploiting data, but an essential layer. Institutions ignore privacy at their peril, and the perils perhaps greatest for universities as guardians of public trust. Technologies tend to advance at a much faster pace than does the law or social practice.[96] When the technologies are in the realm of ideas and knowledge production, as is the case with research and grey data, the stakes for universities are especially high.

### A.    ACCESS TO DATA

Determining who has access to what data, by what criteria, when, and under what conditions is an overarching problem of data governance and stewardship. Competing values are often at stake. Openness promotes transparency and accountability, but can undermine privacy, confidentiality, and anonymity. Confidentiality is essential to protecting human subjects but can limit the uses of data and the ability to reuse data for other purposes. Trust derives from openness in some situations and confidentiality in others. Long-term stewardship is necessary for longitudinal research and for many kinds of data aggregation but may result in retaining sensitive records that should be purged regularly by law, policy, or ethical judgment. Access policies that apply to any given data collection may be multiple, conflicting, and change over time.

Privacy concerns abound at the intersection of research and grey data due to the vagaries of defining "research" and "research data." As discussed above, the boundaries of what is considered research are fluid. Materials collected for administrative or teaching purposes may later be considered useful for research. Conversely, data collected for research purposes might be put to practical use in university operations later.

---

96. *See generally, e.g.*, LAWRENCE LESSIG, CODE AND OTHER LAWS OF CYBERSPACE (1999) (arguing that cyberspace can be regulated by norms, markets, and technological architecture where law fails to keep pace with advancements); LAWRENCE LESSIG, THE FUTURE OF IDEAS: THE FATE OF THE COMMONS IN A CONNECTED WORLD (2001) (explaining how the internet revolution has produced a counterrevolution of creativity and how the legal landscape protected this free space); DANIEL J. SOLOVE & PAUL M. SCHWARTZ, PRIVACY LAW FUNDAMENTALS (2017); SOLOVE, *supra* note 58; JULIE E. COHEN, CONFIGURING THE NETWORKED SELF: LAW, CODE, AND THE PLAY OF EVERYDAY PRACTICE (2012); ANITA ALLEN & MARC ROTENBERG, PRIVACY LAW AND SOCIETY (3d ed. 2015).

One of the major difficulties in implementing policies for open access to research data is the lack of agreement on what content, formats, media, or artifacts are subject to release. Funding agencies and journals generally leave these specifics to investigators, as data may be released in varying states of processing. Rules and practices vary widely by agency and research domain. "Raw" data may be released, with or without sufficient documentation to make them useful to others. Conversely, highly processed data might be released, with or without sufficient documentation, software, and code to make them useful to others. Investigators may meet "the letter of the law" by releasing enough information to satisfy agency or journal requirements, while retaining control over proprietary materials that assure a competitive edge in research. Privacy protection may or may not be an issue, depending on the content of the data.[97]

When disputes arise between researchers, collaborators, funding agencies, or journals about what data are subject to release, universities may need to arbitrate in this unsettled territory. Particularly sensitive, for example, are data from grant projects that constitute dissertation research. To ensure that students can complete their degrees, that research subjects' confidentiality is protected, and that grant contracts are completed, balancing tests may be necessary. Among the reasons that research data are not released is that specific responsibility for depositing or posting data may be unclear. In most domains, data release is not part of regular scholarly practice. Rarely are the principles or mechanics of data management and dissemination covered in graduate courses on research methods. Graduate students and post-doctoral fellows are the primary data-handlers in most research teams. They may not

---

97. *See generally* Christine L. Borgman et al., *Knowledge Infrastructures in Science: Data, Diversity, and Digital Libraries*, 16 INT'L J. ON DIGITAL LIBR. 207 (2015) (discussing the need for expertise in digital libraries, data science, and data stewardship); Christine L. Borgman, Jillian C. Wallis & Matthew S. Mayernik, *Who's Got the Data? Interdependencies in Science and Technology Collaborations*, 21 COMPUTER SUPPORTED COOPERATIVE WORK 485 (2012) (reporting on a long-term study of collaboration between environmental scientists, computer scientists, and engineering research teams as part of a five-university distributed science and technology research center devoted to embedded networked sensing); "RAW DATA" IS AN OXYMORON, *supra* note 89 (arguing that data are not natural resources, but rather cultural ones that need to be generated, protected, and interpreted); Pasquetto, Randles & Borgman, *supra* note 38; Wallis, *supra* note 81 (suggesting that including author contribution statements in publications would assist future users of those data in determining the appropriate contact person for questions about their creation and context).

take, or be given, the data deposit responsibility by principal investigators, for example.[98]

Despite elaborate rules about what constitutes human subjects research, IRBs vary in their judgment of how sensitive any given study may be. For example, IRBs may disagree about necessary protections for records of online activity or historical records. A recent study conducted by researchers at Cornell University and Facebook that manipulated Facebook feeds raised a firestorm of ethical issues in mainstream and social media. A central question raised was when, and to what degree, did the university's IRB review the proposal. The study appears to be legal, per Facebook user agreements; experts disagree about the ethics of using information about individuals in this way.[99]

If an IRB decides that a project does not require IRB review, investigators and staff may have no alternative venue to consult. If sensitive data collection originates outside of the research realm, such as learning analytics, no consultation source may exist beyond the boundaries of the system or project. Only if and when someone wishes to publish findings from such studies does an IRB review them, by which time sensitive data may have been collected inappropriately. These data could fall under open access release policies, depending on funding sources and publication venues. UCLA is unusual in providing an alternative consulting entity, which is the Privacy and Data Protection Board. That board is advisory and consists of faculty and administrators with a broad array of expertise in privacy matters.[100] The board has considered topics such as the content of administrative surveys of campus climate, requests to monitor online activity on campus networks, and policies for surveillance cameras and drones. Other UC campuses established similar boards as part of implementing the recommendations of the UC Privacy and Information Security Initiative.[101] The UC Academic Computing and Communications Committee is assessing ways to harmonize information technology and data governance across the UC campuses.[102]

---

98. *See* Wallis, *supra* note 81; *see also* Jillian C. Wallis, Elizabeth Rolando & Christine L. Borgman, *If We Share Data, Will Anyone Use Them? Data Sharing and Reuse in the Long Tail of Science and Technology*, 8 PLOS ONE e67332, e67332 (2013) (showing that releasing, sharing, and reusing data in CENS reaffirms "the gift culture of scholarship, in which goods are bartered between trusted colleagues rather than treated as commodities").

99. *See* Reed Albergotti & Elizabeth Dwoskin, *Facebook Study Sparks Soul-Searching and Ethical Questions*, WALL ST. J. (June 30, 2014, 9:01 PM), www.wsj.com/articles/facebook-study-sparks-ethical-questions-1404172292 [https://perma.cc/WCL5-BJC2] (detailing how Facebook and Cornell manipulated the news feeds of nearly 700,000 Facebook users for a week to gauge whether emotions spread on social media)); *see also* Chris Chambers, *Facebook Fiasco: Was Cornell's Study of 'Emotional Contagion' an Ethics Breach?*, GUARDIAN (July 1, 2014, 2:00 AM), http://www.theguardian.com/science/head-quarters/2014/jul/01/facebook-cornell-study-emotional-contagion-ethics-breach [https://perma.cc/4FB9-CKZM] (arguing that the Facebook and Cornell study violated ethical norms); Adam D. I. Kramer, Jamie E. Guillory & Jeffrey T. Hancock, *Experimental Evidence of Massive-Scale Emotional Contagion Through Social Networks*, 111 PNAS 8788 (2014) (describing results of the large-scale study on the use emotional manipulation); Robinson Meyer, *Everything We Know About Facebook's Secret Mood Manipulation Experiment*, ATLANTIC (June 28, 2014), https://www.theatlantic.com/technology/archive/2014/06/everything-we-know-about-facebooks-secret-mood-manipulation-experiment/373648/ [https://perma.cc/DX3V-NSFK] (outlining the information available from the study and if IRB review occurred); Gail Sullivan, *Cornell Ethics Board Did Not Pre-Approve Facebook Mood Manipulation Study*, WASH. POST (July 1, 2014), www.washingtonpost.com/news/morning-mix/wp/2014/07/01/facebooks-emotional-manipulation-study-was-even-worse-than-you-thought/ [https://perma.cc/UKN6-C3X5] (stating that the Cornell ethics board did not preapprove the Facebook study and that there was international outrage regarding the manipulation without consent).

Technological research that gathers sensitive data is a growing concern, especially when not submitted for IRB review. Researchers in engineering, for example, may have little experience with human subjects research and be unfamiliar with DHHS and IRB rules. When robotics students test image-recognition algorithms by scattering cameras around a campus, they are likely to capture all manner of human activity without notice or consent of the individuals whose images and actions are recorded. Drones are the current technology of concern, due to their surveillance capabilities and potential for harm to persons and property. Universities are beginning to grapple with ways to balance data protection with innovation in these areas.[103] Technical data such as these could be subject to open access policies and could inadvertently be released even though they are subject to PII or other protections.

B.        USES AND MISUSES OF DATA

Among the greatest promises of "big data" is the ability to exploit data for innovative purposes, especially uses that were not anticipated at the time of data collection.[104] Data exploitation can lead to scientific breakthroughs,

---

100. *See About the UCLA Board on Privacy and Data Protection*, UCLA, http://privacyboard.ucla.edu/ [https://perma.cc/GME2-UMY8] (last visited Aug. 15, 2018).

101. *See* UC PRIVACY & INFO. SEC. INITIATIVE STEERING COMM., *supra* note 5.

102. UNIV. OF CAL. ACAD. SENATE, *supra* note 94.

103. *See* Brandon Stark, *UC Unmanned Aircraft System Safety*, UNIV. OF CAL., OFFICE OF THE PRESIDENT, http://www.ucop.edu/enterprise-risk-management/resources/centers-of-excellence/unmanned-aircraft-systems-safety.html [https://perma.cc/RD8W-7C5L] (last visited Aug. 15, 2018) (detailing safety guidance for unmanned aircraft).

104. *See* BORGMAN, *supra* note 6; *see also generally* Tom Kalil, *Big Data is a Big Deal*, WHITE HOUSE (Mar. 29, 2012, 9:23 AM), https://obamawhitehouse.archives.gov/blog /2012/03/29/big-data-big-deal [https://perma.cc/H3LQ-XXBR] ("By improving our ability to extract knowledge and insights from large and complex collections of digital data, the initiative promises to help accelerate the pace of discovery in science and engineering, strengthen our national security, and transform teaching and learning."); ROB KITCHIN, THE DATA REVOLUTION: BIG DATA, OPEN DATA, DATA INFRASTRUCTURES AND THEIR CONSEQUENCES (1 edition ed. 2014) (offering an overview of big data, open data, and data philosophical insights, and to new products and services. When data exist, clever people will find new uses for those data. The challenge is how to encourage innovation while protecting against inappropriate, privacy-invading uses of those data. Data systems subject to strict compliance regulations such as IRB, HIPAA, FERPA, and PII may be a declining portion of university data acquisition. The privacy frontier is the vast territory outside those regulated systems.

### 1. Anticipating Potential Uses and Misuses

When the Code of Fair Information Practices was developed nearly forty years ago, data collection was vastly smaller in scale and information systems were more discrete entities. At today's scale of data collection and aggregation, the original FIPS principles provide much less privacy protection. Revisions of FIPS issued in 2013 by the OECD addressed practical implementations based on risk management and improvements in interoperability of data systems.[105] Notice and informed consent, the foundational FIPS principles, remain necessary but are no longer sufficient.[106] When individuals consent to the collection of specific data elements, they may be giving much broader permissions than anticipated, especially when the stated purposes provide wide latitude for use in research, personalization, improving system performance, or other vagaries.

Broader data collection, for more generic purposes, increases the potential for misuses of data and for privacy risks. The benefits and risks of big data in universities can be balanced by two means. One way is to adhere more broadly to the FIPS principles, including collection limitation, data quality, use specification, and purpose specification principles. Common to both FIPS and the tenets of privacy by design is to limit data collection and to state an express justification for each data element to be acquired.[107] The second means is to

---

infrastructures including analysis of the technical shortcomings of the data revolution and the implications for academic, business, and governemnt practices); PRIVACY, BIG DATA, AND THE PUBLIC GOOD: FRAMEWORKS FOR ENGAGEMENT xi (Julia Lane et al. eds., 1st ed. 2014) ("The book's authors paint an intellectual landscape that includes the legal, economic, and statistical context necessary to frame the many privacy issues [of data] . . . ."); MAYER-SCHÖNBERGER & CUKIER, *supra* note 8.

105. *See* THE OECD PRIVACY FRAMEWORK, *supra* note 60.

106. Susan Landau, *Control Use of Data to Protect Privacy*, 347 SCIENCE 504–06 (2015); UCLA DATA GOVERNANCE TASK FORCE, *supra* note 5; SOLOVE, *supra* note 58.

107. Philip E. Agre, *Institutional Circuitry: Thinking About the Forms and Uses of Information*, 14 INFO. TECH. & LIBR. 225 (1995); Bellotti & Sellen, *supra* note 9; ANN CAVOUKIAN, PRIVACY BY DESIGN: THE 7 FOUNDATIONAL PRINCIPLES (2011), https://www.ipc.on.ca/wp-content/uploads/Resources/7foundationalprinciples.pdf [https://perma.cc/4RGR-FPUY] (arguing that mere compliance with regulatory frameworks is insufficient to protect privacy, govern uses of data once collected. Governance should include specifying who has access to what data, when, and under what circumstances, and identifying what uses are considered appropriate and inappropriate. As criteria for these judgments can change over time, governance processes to assure continuing oversight also are needed.[108]

Individual data elements that appear innocuous at the time of collection can become sensitive in later contexts. In a recent example, students' permanent addresses, which universities maintain in case of emergency, may reveal legal residency status to immigration authorities. Recent changes in the status of "Dreamers" (undocumented students who were brought to the United States as minors) made this information extremely sensitive.[109] Similarly, most universities provide minimal information about students in public directories out of concern for stalking and other harms.

Potential misuse of research data is a concern often mentioned by those reluctant to release data associated with grants or publications.[110] Data can be taken out of context to make misleading or incorrect inferences, as when health and climate data are used selectively to make claims that run counter to those of the investigators.[111]

### 2. Reusing Data

One person's good use or reuse of data may be seen by others as a misuse. The ability to reuse data effectively depends on factors such as the quality of the original data collection, the degree of documentation provided to interpret protocols and context, and the availability of associated software, code, and instrumentation.[112] Whether research data or grey data, problems arise in

---

and that organizations should instead implement privacy-protective features as a default element); OECD PRIVACY FRAMEWORK, *supra* note 60.

108. *See* UCLA DATA GOVERNANCE TASK FORCE, *supra* note 5; *see also* UC PRIVACY & INFO. SEC. INITIATIVE STEERING COMM., *supra* note 3.

109. *See* Adam Harris, *Colleges Deplore Trump's Threat to DACA. How Far Can They Go to Fight It?*, CHRON. HIGHER EDUC. (Sept. 6, 2017), http://www.chronicle.com/article/ Colleges-Deplore-Trump-s/241110 (demonstrating how colleges can be "safe zones" for undocumented students).

110. Wallis, Rolando & Borgman, *supra* note 98; *see also* BORGMAN, *supra* note 6.

111. *See generally* Paul N. Edwards, *Global Climate Science, Uncertainty and Politics: Data-Laden Models, Model-Filtered Data*, 8 SCI. AS CULTURE 437 (1999) (examining how data is important in legitimizing political activity around global climate change); PAUL N. EDWARDS, A VAST MACHINE: COMPUTER MODELS, CLIMATE DATA, AND THE POLITICS OF GLOBAL WARMING (2010) (tracing the history of efforts to gather weather and climate records for the whole planet and the resulting "data friction"); BEN GOLDACRE, BAD SCIENCE: QUACKS, HACKS, AND BIG PHARMA FLACKS (2008); GOLDACRE, *supra* note 84.

112. *See, e.g.*, Pasquetto, Randles & Borgman, *supra* note 38; Matthew S. Mayernik, *Research Data and Metadata Curation as Institutional Issues*, 67 J. ASS'N FOR INFO. SCI. & TECH. 973 (2016) measurement because collecting good data is hard to do. Considerable sophistication in the design of research or other protocols is necessary, combined with expertise in statistics and methods of data cleaning.[113] Surveys, for example, are far more complex to design, execute, analyze, and interpret than is apparent to the novice researcher—or to the staff member assigned to evaluate a service or system. Problems also arise in interpreting and drawing inferences from data because much must be known about the purposes and context in which the data were collected.

The potential for misuse and abuse multiply when data elements are aggregated, whether from one data resource or many. Variable names, units of measurement, research protocols, and circumstances of data collection introduce errors that are difficult to assess when combining data. Reliability and validity concerns abound. Estimates of the amount of labor required to "clean" data for aggregation are hard to find; one source suggests devoting about eighty percent of the work to cleaning and integration.[114] Data science is an inexact science, at best.

Despite these cleaning and analysis problems, data scientists have been remarkably effective at reidentifying individuals by aggregating records from multiple sources.[115] Researchers who wish to use sensitive data about individuals, such as medical records or certain types of surveys, often are required to sign agreements that they will not attempt to re-identify the research subjects.[116]

Intellectual property concerns also arise in aggregating data from multiple sources, whether from research, administrative, or external sources. Although any individual dataset may carry documentation about ownership and

---

(examining variability in data and metadata practices using "institutions" as the key theoretical concept); BORGMAN, *supra* note 6.

113.  *See* Kreuter & Peng, *supra* note 46, at 267–69 (describing the statistical changes of integrating big data); *see also* WILLIAM R SHADISH, THOMAS D. COOK & DONALD T. CAMPBELL, EXPERIMENTAL AND QUASI-EXPERIMENTAL DESIGNS FOR GENERALIZED CAUSAL INFERENCE (2002).

114.  MAYER-SCHÖNBERGER & CUKIER, *supra* note 8.

115.  *See, e.g.*, Boris Lubarsky, *Re-Identification of "Anonymized Data"*, 1 GEO. L. TECH. REV. 202, 211–12 (2017); *see also* Yves-Alexandre de Montjoye et al., *Unique in the Shopping Mall: On the Reidentifiability of Credit Card Metadata*, 347 SCIENCE 536 (2015); *see also* Latanya Sweeney, *k-Anonymity: A Model for Protecting Privacy*, 10 INT'L J. ON UNCERTAINTY FUZZINESS & KNOWLEDGE-BASED SYS. 557 (2002) (suggesting methods of keeping released data ambiguous to prevent reidentification); Latanya Sweeney, Matching Known Patients to Health Records in Washington State Data (July 5, 2013) (unpublished manuscript), https://arxiv.org/abs/1307.1370 [https://perma.cc/N8J9-JSK3].

116.  Jared A. Lyle, George C. Alter & Ann Green, *Partnering to Curate and Archive Social Science Data*, *in* RESEARCH DATA MANAGEMENT: PRACTICAL STRATEGIES FOR INFORMATION PROFESSIONALS (2014); NAT'L RESEARCH COUNCIL, *supra* note 65.

licensing, maintaining intellectual property information in provenance records through multiple generations of use is proving to be a frontier problem in the data sciences. Despite attaching licenses to datasets that protect privacy, that information can be lost downstream.[117]

### 3. Responsibilities for Data Collections

Responsibility for data is particularly diffuse in universities, although similar issues arise in all institutions. Research data collections are scattered across labs and stored on laptops or local servers. Multiple generations of students and staff may have access to these data, which can cumulate over long periods of time. Few of these data may involve human subjects and few of these data may be privacy-sensitive, especially when used alone. Similarly, vast collections of grey data are scattered across universities and cumulated over time. Many are purged regularly on a records-retention cycle, but many are not. Access to campus collections may be limited to the few staff who are certified for their use. In other cases, generations of student workers and other transient labor may use grey data daily in their jobs.

As universities outsource more computing systems and services to commercial entities, they relinquish a substantial degree of control over the data collected by their online systems. When universities purchase licenses for access to digital resources such as publications and grey literature, those contracts may allow data providers to track usage by identifiable individuals, in ways that undermine libraries' abilities to protect traditional rights to read anonymously.[118] Similar problems arise when universities partner with vendors for shared usage of data about individuals, such as analytics on learners or patients, whether for graduation rates or treatment outcomes. Universities are becoming more sophisticated about building privacy and security protections into contracts, especially where vendors have offered to sell universities data about their users.

---

117. Chaitanya Baru, *Sharing and Caring of eScience Data*, 7 INT'L J. DIGITAL LIBR. 113 (2007); Jane Hunter & Kwok Cheung, *Provenance Explorer-A Graphical Interface for Constructing Scientific Publication Packages from Provenance Trails*, 7 INT'L J. DIGITAL LIBR. 99 (2007); Mayernik, *supra* note 112; Andrew E. Treloar & Mingfang Wu, *Provenance in Support of the ANDS Four Transformations*, 11 INT'L J. DIGITAL CURATION 183 (2016); Michael Wright et al., *Connecting Digital Libraries to eScience: The Future of Scientific Scholarship*, 7 INT'L J. DIGITAL LIBR. 1 (2007); Paul Groth et al., *Requirements for Provenance on the Web*, 7 INT'L J. DIGITAL CURATION 39 (2012); James Cheny et al., *Requirements for Provenance on the Web*, W3C PROVENANCE INCUBATOR GROUP (Dec. 7, 2010, 11:52 PM), http://www.w3.org/2005/Incubator/prov/wiki/User_Requirements [https://perma.cc/2BH8-6MPL].

118. Lynch, *supra* note 83; Cohen, *supra* note 83; *An Interpretation of the Library Bill of Rights*, AM. LIBRARY ASS'N (July 1, 2014), http://www.ala.org/advocacy/intfreedom/librarybill/interpretations/privacy [http://perma.cc/Y8Z5-WBKG].

Yet harder problems arise when faculty or staff require students to use third-party online tools that are not licensed by the university. These "free" online tools are attractive because they offer sophisticated activities, content, or evaluation capabilities suitable for a particular course. However, these tools collect personal data about their users that are shared with outside partners, barring contracts to the contrary. Students may have little choice but to opt-in to usage agreements if the software is required for course activities. A growing concern is liability when such vendors breach confidential student or faculty data, especially when no contract exists between the university and the vendor to ensure protections. Instructors and students too often are unaware of the privacy and security risks such as these. Despite university policies and warnings by technology professionals not to install such software, usage can be difficult to detect, especially by understaffed tech support offices. A shadow network of risky technology lurks on many campuses.

C.       PUBLIC RECORDS REQUESTS

Given the continuing advances toward open access to publications and to data over the last several decades, it is counter-intuitive to place public records requests on the privacy frontier. Public access laws are essential to democratic societies, and university researchers often avail themselves of these laws in gaining access to information.[119] However, these laws are being used in political and frivolous ways that threaten academic freedom and privacy.[120]

Law and policy about university data collections are often ambiguous, which raises two related questions. One is that the more data that universities collect, the larger the pool of resources subject to public records requests. Hence the principle, "if you can't protect it, don't collect it." Research data on controversial topics such as climate change, guns, tobacco, and abortion are among the most common records requests.[121] Releasing data and communications about research in progress threatens academic freedom and

---

autonomy privacy. State public records laws vary in the degree to which they allow exceptions for research material.

Grey data also can be requested, such as information on the demographics of the student body, marital status of individuals in an academic department, or email correspondence of individual faculty or administrators.[122] As public records requests to universities have become more sophisticated, so have the responses of university counsel.[123]

The second issue is that state public records act requests in the United States apply to public universities but not to private universities or corporations. Faculty, students, and staff at public universities thus carry a higher burden in managing their data and in responding to public records requests. Responding to such requests can be extremely time-consuming and expensive, in addition to the risks to academic freedom and privacy. Researchers at public and private universities frequently collaborate with each other, which can expose the data of private universities to these requests. As a result, members of public universities may seek protections of their research materials and communications comparable to those at private universities, which also protects collaborations.[124]

An emerging area of concern is whether trends toward open access to data in some fields may undermine a university's ability to protect data from public records requests in other fields. In some domains of the biosciences, physical sciences, and social sciences, open data is the default condition at the time of publishing research. Some researchers in some domains attempt to work completely in the open, releasing data continuously. In most academic

disciplines, however, researchers maintain control of their data and records indefinitely.[125]

## D.     CYBER RISK AND DATA BREACHES

Universities are the third highest sector for data breaches, constituting about ten percent of reported breaches; healthcare and retail are the top two sectors.[126] From 2005 to late-2017,[127] colleges and universities reported about 800 breaches, affecting more than twenty-five million records.[128] Institutions of higher education have extensive data resources and may be perceived as more vulnerable to attack than hospitals, banks, governments, retail, or other entities. Research universities are commonly targeted for the intellectual property manifest in research content. Those with medical centers are targeted for patient records, which are valuable resources for identity and insurance theft. Student records have become high value targets because logon credentials provide access to expensive licensed content from publishers and other sources. Intruders seeking one kind of information may wander through other databases along the way. Data on individuals that are held by third parties, such as collaborating universities or outside contractors, also are vulnerable to breach.

Education is the institutional sector facing the greatest challenges in balancing access and protection. Universities are heterogeneous institutions that acquire many kinds of data and need sophisticated, layered approaches to cyber security. Whereas the financial and intelligence sectors, for example, may prioritize cyber risk protection in the extreme, universities are open by design, encouraging the free flow of information throughout their communities. Individuals partner with collaborators from other institutions, countries, and cultures, which requires shared access to online resources. Campus visitors are vast in number and need access to networks to participate in local activities. Student and staff turnover is high due to short courses and short-term contracts. As a result of these operating conditions, universities must secure their systems and networks without crippling their missions of research, teaching, and service. Research data must flow to students for use in class projects, albeit in a controlled manner. Network security must not become ubiquitous with surveillance.

---

125. BORGMAN, *supra* note 6, at 276.

126. SYMANTEC, ISTR20: INTERNET SECURITY THREAT REPORT 17 (2015), https://www.symantec.com/content/en/us/enterprise/other_resources/21347933_GA_R PT-internet-security-threat-report-volume-20-2015.pdf [https://perma.cc/2WHW-DVHX].

127. For purposes of this Article, the time period of analysis extends to November 3, 2017.

128. *Data Breaches*, PRIVACY RIGHTS CLEARINGHOUSE, https://www.privacyrights.org/data-breaches [http://perma.cc/DG2S-E8Z8] (last visited Aug. 15, 2018).

Cyber risk takes many forms, such as phishing attacks on individuals, viruses, bots, ransomware, data breaches, and distributed denial of service attacks. The list grows by the day. Some risks are obvious, such as the need for many layers of protection on patient data. Others are less obvious, such as attacking a student admissions database for competitive information. Systems are only as well protected as their weakest link. The Target Store breach of credit card records resulted from a successful hack of their HVAC system.[129] A distributed denial of service attack on Netflix was launched by mobilizing networked household devices, most notably baby monitors.[130] The ability to mobilize small devices for big attacks will grow as the Internet of Things expands, potentially becoming the "Internet of Terror."[131]

Universities are following the lead of the public and private sectors in enhancing security of their systems, training their communities, and promoting good practices for "cyber health." Deleterious computer-related events are difficult to anticipate, and no sector of the economy is immune to attack.[132] No online system ever can be completely secure, any more than any building is completely secure from physical attack. By analogy, security comes in layers of locks, cameras, sensors, and alerts.[133] Resilience and recovery also have become watchwords for cybersecurity. The severity of attacks must be minimized, but backup and recovery plans also are necessary.[134] The costs and benefits of each tactic must be evaluated, lest funds spent on protection lessen the investment in the mission of the institution.

---

129. Jaikumar Vijayan, *Target Attack Shows Danger of Remotely Accessible HVAC Systems*, COMPUTERWORLD (Feb. 7, 2014, 6:52 AM), www.computerworld.com/article/2487452/cybercrime-hacking/target-attack-shows-danger-of-remotely-accessible-hvac-systems.html [http://perma.cc/GSE3-GXD3].

130. Haley Sweetland Edwards, *How Web Cams Helped Bring Down the Internet, Briefly*, TIME (Oct. 25, 2016), http://time.com/4542600/internet-outage-web-cams-hackers/ [http://perma.cc/Z6XX-RVHL].

131. George V. Neville-Neil, *IoT: The Internet of Terror*, 60 COMM. ACM 36 (2017).

132. *See* Peter G. Neumann, *Far-Sighted Thinking About Deleterious Computer-Related Events*, 58 COMM. ACM 30 (2015); Taylor Armerding, *The 17 Biggest Data Breaches of the 21st Century*, CSO ONLINE (Jan. 26, 2018, 3:44 AM), https://www.csoonline.com/article/2130877/data-breach/the-16-biggest-data-breaches-of-the-21st-century.html [http://perma.cc/39QF-EWBP]; Waqas Amir, *Unprotected S3 Cloud Bucket Exposed 100GB of Classified NSA Data*, HACKREAD, (Nov. 29, 2017), https://www.hackread.com/unprotected-s3-cloud-bucket-exposed-100gb-of-classified-nsa-data/ [http://perma.cc/8FYA-8DVJ] (describing a breach that made public classified information for political leaders and U.S. military); David Greene, *NSA's Hackers Were Themselves Hacked in Major Cybersecurity Breach*, NPR (Nov. 14, 2017, 5:00 AM), https://www.npr.org/2017/11/14/564006460/nsas-hackers-are-hacked-in-major-cybersecurity-breach [http://perma.cc/9YVF-2UF6]; Andy Greenberg, *He Perfected a Password-Hacking Tool—Then the Russians Came Calling*, WIRED (Nov. 9, 2017, 7:00 AM), https://www.wired.com/story/how-mimikatz-became-go-to-hacker-tool/ [http://perma.cc/N6XW-QU7C] (describing a Russian hacker attempting to steal a French programmer's source code in a hotel lobby in Moscow); Julie Angwin, *How Journalists Fought Back Against Crippling Email Bombs*, WIRED (Nov. 9, 2017, 7:00 AM), https://www.wired.com/story/how-journalists-fought-back-against-crippling-email-bombs/ [http://perma.cc/SV3J-8PF5]; Susan Landau, *The Real Security Issues of the iPhone Case*, 352 SCIENCE 1398 (2016).

A looming challenge on the privacy frontier is how to secure the privacy of human subjects once data are collected. IRBs focus on the design of studies, confidentiality, notice and consent, and good practices for data storage and backup. Their membership is drawn from researchers across campus who have expertise in research design and methods. IRBs, and the university staff that support them, are not necessarily experts in security, cyber risk, cryptography, or in the open data policies to which research projects may be subject. Investigators are required to report on research progress at regular intervals. However, short of known data breaches, IRBs have few mechanisms to follow up on data security. Data management practices vary widely by domain, thus IRBs lack common standards to enforce across campuses.[135] Governance models need to promote more engagement between IRBs, investigators, cyber security units, and other parts of the research enterprise. Among the concerns that universities and other sectors must address are methods of anonymization; responsibilities for data protection, release, and stewardship; and accountability for secure and responsible data management practices.

E.      CURATING DATA FOR PRIVACY PROTECTION

Data management is an expensive endeavor, and one that has come to the fore in the research data arena.[136] Any entity that collects data must make

---

conscious decisions about which data are worth sustaining, which can be discarded, and which might be allowed to fade away.[137] Maintaining privacy protections and reducing risks is essential to accomplishing these goals.

Digital data do not survive by benign neglect. Continuous investments are required to refresh computers, storage devices, software, and websites. Regular technology maintenance is but a baseline for longer term data curation, however. Larger challenges arise when software is updated, is no longer available, or is not supported; when computer ports and drivers are not compatible with current equipment; when data processing pipelines are poorly documented; and when those with critical expertise graduate or leave the university. Thus, digital data remain useful only through investments in curation, documentation, and migration to new formats and systems. Systems and data collections need to be assessed on a cyclical basis, purging sensitive data based on retention rules and refreshing data collections worthy of continuing access. Maintaining provenance records is essential, lest data collections be separated from information about origins; licensing and ownership; applicable regulations; records of notice, consent, and acceptable uses; authorizations for access; and other contexts.[138] Archivists, records managers, and librarians should be closely involved in these processes.[139]

Responsibility for data collections is highly distributed in universities, which complicates curating data collections in the short and long term. A researcher with expertise in data management or with the resources to invest in long-term sustainability of research data is rare. Even more rare is the researcher with expertise in data archiving, records management, and the legal vagaries of records retention cycles. Similarly, few of the administrative staff involved in collecting and analyzing grey data are records management experts.

---

137. Christine L. Borgman, *Not Fade Away: Social Science Research in the Digital Era*, PARAMETERS (June 23, 2016), http://parameters.ssrc.org/2016/06/not-fade-away-social-science-research-in-the-digital-era/ [https://perma.cc/8JSQ-YDE9].

138. Miriam Ney, Guy K. Kloss & Andreas Schreiber, *Using Provenance to Support Good Laboratory Practice in Grid Environments*, *in* DATA PROVENANCE AND DATA MANAGEMENT IN ESCIENCE 157, 157–59 (Qing Liu et al. eds., 2011); Lucian Carata et al., *A Primer on Provenance*, 57 COMM. ACM 52, 52 (2014) ("[D]iscussing not only existing systems and the fundamental concepts needed for using them in applications today, but also future challenges and opportunities."); Jinfang Niu, *Provenance: Crossing Boundaries*, 41 ARCHIVES & MANUSCRIPTS 105 (2013) (surveying different provenance practices for different subject matters, data types, and professional field specializations); PROVENANCE AND ANNOTATION OF DATA (Ian Foster & Luc Moreau eds., 2006); Clifford A. Lynch, *When Documents Deceive: Trust and Provenance as New Factors for Information Retrieval in a Tangled Web*, 52 J. AM. SOC'Y FOR INFO. SCIENCE & TECH. 12 (2001); Jun Zhao et al., *Linked Data and Provenance in Biological Data Webs*, 10 BRIEFINGS IN BIOINFORMATICS 139 (2009).

139. Ney, Kloss & Schreiber, *supra* note 138; Carata et al., *supra* note 138, at 52–60; Niu, *supra* note 138; FOSTER & MOREAU, *supra* note 138; Lynch, *supra* note 138; Zhao et al., *supra* note 138.

All of these individuals and offices need somewhere to turn for guidance and responsibility to ensure that universities make wise choices for what to keep, what to discard, how, and when.

Institutions more readily claim ownership of data than take responsibility for curating those data. Ownership and stewardship need to be more tightly coupled in universities, and probably in most other types of institutions.

## V.      CONCLUSIONS AND RECOMMENDATIONS

Universities are as enamored of "big data" as other sectors of the economy and are similarly effective in exploiting those data to competitive advantage. They have privileged access to research data and to data about their communities, all of which can be mined and combined in innovative ways. Universities also have a privileged social status as guardians of the public trust, which carries additional responsibilities in protecting privacy, academic and intellectual freedom, and intellectual property. They must be good stewards of the data entrusted to them, especially when conflicts arise between community practices and values. For some kinds of data, good stewardship requires that access to data be sustained indefinitely, and in ways that those data can be reused for new purposes. For other kinds of data, good stewardship requires that they be protected securely for limited periods of time and then destroyed. Factors that distinguish data worth keeping or discarding vary widely by domain, content, format, funding source, potential for reuse, and other circumstances.[140] Criteria for data protection and access also can change over time, whether due to different uses of a data collection, such as grey data being mined for research or research data being deployed for operations; transfer of stewardship within and between institutions; changes in laws and policies; or new externalities.

The rate of data collection has grown exponentially over the last decade through both research and grey data within universities, along with data collection in the other economic sectors with which universities partner. These include government and business, social media, sensor networks, the Internet of Things, and much more. As the ability to mine and combine data improves, and technologies become more interoperable, the boundaries between data types and origins continue to blur. Responsibilities for stewardship and exposure to cyber risk increases accordingly. Risks to privacy invasion, both information privacy and autonomy privacy, accelerate as most of these data can be associated with individuals, whether as content or creators of data. Anonymity, which is fundamental to most methods of privacy protection, has become extremely difficult to sustain as methods of re-identifying individuals become more sophisticated. Notice and informed consent remain necessary but are far from sufficient for maintaining privacy in universities or in other sectors.

---

140.   BORGMAN, *supra* note 6, at 271–87.

Open access to publications and to data are social policies that promote transparency and accountability in the research enterprise. Adoption is uneven because costs, benefits, and incentives for open access, especially to data, are aligned in only a few fields and domains. For most researchers, releasing data involves considerable costs, with benefits going to others. These costs may include curation (e.g., providing metadata, documentation, and records of provenance and licensing), computer storage and maintenance, software acquisition and maintenance, migration to new software and hardware, and fees for data deposit. Disposal of data also involves costs to assess what to keep and what to discard, and to ensure safe destruction of confidential or proprietary materials. Individual researchers, their employers, or their funders may bear the costs of data stewardship and responsibilities for protecting privacy, academic and intellectual freedom, intellectual property, and other values.

None of these frontier challenges is easily addressed, nor will appropriate responses be consistent across the university sector in the U.S., much less in other countries and cultures. Data are valuable institutional assets, but they come at a price. Individuals and institutions must be prepared to protect the data they collect. These recommendations, which draw heavily on experiences in the University of California, are offered as starting points for discussion.

A.        BEGIN WITH FIRST PRINCIPLES

Universities should focus on their core missions of teaching, research, and services to address priorities for data collection and stewardship. Tenets of privacy by design, the Code of Fair Information Practice, the Belmont Report,[141] and codifications of academic and intellectual freedom are established and tested. Implementation is often incomplete, however. For faculty, students, staff, research subjects, patients, and other members of the university community to enjoy protection of information and autonomy privacy, more comprehensive enforcement of principles such as limiting data collection, ensuring data quality, and constraining the purposes for each data element is necessary. Digital data do not survive by benign neglect, nor are records destroyed by benign neglect. Active management is necessary. Notice and consent should never be implicit. When institutions ask for permission to acquire personal data, are transparent, and are accountable for uses of data, they are more likely to gain respect in the court of public opinion.

---

141. BELMONT REPORT, *see supra* note 64, provides the basis for human subjects regulation by Institutional Review Boards, as governed by the DHHS. *See supra* Section II.A.1.

B.        EMBED THE ETHIC

Data practices, privacy, academic and intellectual freedom, intellectual property, trust, and stewardship all are moving targets. Principles live longer than do the practices necessary to implement those principles. Universities are embedding data science and computational thinking into their curricula at all levels. This is an opportune moment to embed data management, privacy, and information security into teaching and practice as well. By encouraging each individual to focus on uses of data, the problem becomes personal. Rather than collecting all data that could conceivably be collected, and exploiting those data in all conceivable ways, encourage people to take a reflective step backwards. Consider the consequences of data collection about yourself and others, and how those data could be used independently or when aggregated with other data, now and far into the future. Think about potential opportunities and risks, for whom, and for how long. Study data management processes at all levels and develop best practices. Collect data that matter, not just data that are easy to gather. Interesting conversations should ensue. The Golden Rule still rules.

C.        PROMOTE JOINT GOVERNANCE

The successes of the University of California in developing effective principles for governing privacy and information security have resulted from extensive deliberations between faculty, administrators, and students. These can be long and arduous conversations to reach consensus but have proven constructive at creating communication channels and building trust. Many years of conversations about information technology policy at UCLA, for example, have resulted in much deeper understanding between parties. Faculty have learned to appreciate the challenges faced by administrators who need to balance competing interests, keep systems running, and pay for infrastructure out of fluctuating annual budgets. Administrators, in turn, have learned to appreciate the challenges faced by faculty who have obligations to collaborators, funding agencies, and other partners scattered around the world, and daily obligations to support students who have disparate skills and access to disparate technologies. Institutional learning is passed down through generations of faculty, students, and administrators through joint governance processes. These mechanisms are far from perfect and can be slow to respond at the pace of technological change. However, echoing Churchill's assessment of democracy, it works better than any other system attempted to date.

D.    PROMOTE AWARENESS AND TRANSPARENCY

The massive data breaches of Equifax, Target stores, J.P. Morgan Chase, Yahoo, the National Security Agency, and others have raised community awareness about data tracking, uses of those data by third parties, and the potential for exposure.[142] This is an ideal time to get the community's attention about opportunities and risks inherent in data of all kinds. Individuals, as well as institutions, need to learn how to protect themselves and where to place trust online. People may react in anger if they suspect that personal data are being collected without notice and consent or think they are being surveilled without their knowledge.[143] Universities are at no less cyber risk than other sectors but are still held to higher standards for the public trust. They have much to lose when that trust is undermined.

E.    DO NOT PANIC

Panic makes people risk-averse, which is counterproductive. Locking down all data lest they be released under open access regulations, public records requests, or breaches will block innovation and the ability to make good use of research data or grey data. The opportunities in exploiting data are only now becoming understood. Balanced approaches to innovation, privacy, academic and intellectual freedom, and intellectual property are in short supply. Patience and broad consultation of stakeholders is needed.

---

142.   *See* Armerding, *supra* note 132; Amir, *supra* note 132; Greene, *supra* note 132.

143.   Steve Lohr, *At Berkeley, a New Digital Privacy Protest*, N.Y. TIMES (Feb. 1, 2016), https://www.nytimes.com/2016/02/02/technology/at-uc-berkeley-a-new-digital-privacy-protest.html [https://perma.cc/F6CN-CAEG]; The Associated Press, *Online Attacks at UCLA Health Exposed 4.5 Million*, N.Y. TIMES (July 17, 2015), https://www.nytimes.com/2015/07/18/business/online-attacks-at-ucla-health-exposed-4-5-million.html [http://perma.cc/WE56-LB4Y].